\documentclass[prd,aps,preprint,nofootinbib,amssymb,eqsecnum]{revtex4}

\usepackage{amsmath}
\usepackage{mathrsfs}
\usepackage{slashed}
\usepackage{graphicx}
\usepackage{xcolor}
\usepackage{dsfont}
\usepackage{ulem}

\usepackage[colorlinks,
            linkcolor=black,
            anchorcolor=black,
            citecolor=black
            ]{hyperref}

\newcommand{\GeV}{{\, \rm GeV}}
\newcommand{\MeV}{{\,\rm MeV}}
\newcommand{\fb}{{\,\rm fb}}

\newcommand{\SM}{\text{SM}}
\newcommand{\NP}{\text{NP}}
\newcommand{\mO}{\mathcal O}
\newcommand{\mB}{\mathcal B}
\newcommand{\sfA}{\textsf{A}}
\newcommand{\sfB}{\textsf{B}}

\newcommand{\eu}{\epsilon^u}
\newcommand{\ed}{\epsilon^d}
\newcommand{\el}{\epsilon^\ell}

\newcommand{\heu}{\hat{\epsilon}^u}
\newcommand{\hed}{\hat{\epsilon}^d}
\newcommand{\hel}{\hat{\epsilon}^\ell}

\begin{document}

\baselineskip=15pt

\preprint{NCTS-PH/1810}

\title{Flavor Violating Higgs Couplings in Minimal Flavor Violation}

\author{Jin-Jun Zhang${}^{1}$\footnote{zhangjinjun@sxnu.edu.cn}}
\author{Min He${}^{2}$\footnote{hemind@sjtu.edu.cn}}
\author{Xiao-Gang He${}^{1,2,3,4}$\footnote{hexg@phys.ntu.edu.tw}}
\author{Xing-Bo Yuan$^{3}$\footnote{xbyuan@cts.nthu.edu.tw}}

\affiliation{${}^{1}$School of Physics and Information Engineering, Shanxi Normal University, Linfen 041004, China}
\affiliation{${}^{2}$Tsung-Dao Lee Institute, and School of Physics and Astronomy, Shanghai Jiao Tong University, Shanghai 200240, China}
\affiliation{${}^{3}$Department of Physics, National Taiwan University, Taipei 10617, Taiwan}
\affiliation{${}^{4}$Physics Division, National Center for Theoretical Sciences, Hsinchu 30013, Taiwan}

\begin{abstract}
Motivated by the rencent LHC data on the lepton-flavor violating (LFV) decays $h\to \ell_1 \ell_2$ and $B_{s,d}\to \ell_1 \ell_2$, we study the Higgs-mediated flavor-changing neutral current (FCNC) interactions in the effective field theory (EFT) approach without and with the minimal flavor violation (MFV) hypothesis, and concentrate on the later. After considering the $B$ and $K$ physics data, the various LFV processes, and the LHC Higgs data, severe constraints on the Higgs FCNC couplings are derived, which are dominated by the LHC Higgs data, the $B_s - \bar B_s$ mixing, and the $\mu \to e \gamma$ decay. In the general case and the MFV framework, allowed ranges of various observables are obtained, such as $\mB(B_s\to \ell_1 \ell_2)$, $\mB(h\to \ell_1 \ell_2)$, $\mB(h\to q_1 q_2)$, and the branching ratio of $\mu \to e$ conversion in Al. Future prospects of searching for the Higgs FCNC interactions at the low-energy experiments and the LHC are discussed.
\end{abstract}

\maketitle

\section{Introduction}
The Higgs boson has been discovered at the LHC~\cite{Aad:2012tfa,Chatrchyan:2012xdj}, with a mass of $125\GeV$ and properties consistent with the standard model (SM) predictions. Precision measurements on the Higgs couplings with the SM particles will be one of the most important tasks for the LHC Run II and its high-luminosity upgrade. Any deviation from the SM expectations in Higgs phenomenology is an unambiguous evidence for new physics (NP)~\cite{Csaki:2015hcd,Mariotti:2016owy}.

If there is NP beyond the SM (BSM), the Higgs boson generally can deviate from those predicted in the SM having new flavor-conserving and flavor-changing neutral current (FCNC) interactions. The FCNC Yukawa couplings of Higgs to SM fermions can affect various low-energy precision measurements. In the SM, the FCNC Yukawa interactions are forbidden at the tree level. However, the Higgs-mediated FCNCs generally appear at the tree level in models beyond the SM~\cite{Branco:2011iw,Chiang:2009kb,He:2002ha,Crivellin:2013wna,Kim:2015zla}. These Higgs-mediated couplings can generate the processes which are forbidden in the SM, or enhance some rare decays. In this respect, the lepton-flavor violating (LFV) decays provide excellent probes for such FCNC interactions, such as the $\mu \to e \gamma$, $B_{s,d} \to \ell_1 \ell_2$ and $h \to \ell_1 \ell_2$ decays ($\ell_{1,2} = e,\,\mu,\,\tau$) and can be probed by the LHC and other low energy experiments. 

Recently, significant progresses on searching for such interactions are made at the LHC. Based on $3\fb^{-1}$ data at Run I, a search for the LFV $B_{s,d}$ decays at the LHCb experiment obtains the following upper limits~\cite{Aaij:2017cza}
\begin{align}
  \mB(B_d \to e \mu) < 1.3 \times 10^{-9}\,,
  \qquad
  \mB(B_s \to e \mu) < 6.3 \times 10^{-9}\,,
\end{align}
at 95\% CL. For the LFV Higgs decays, the CMS collaboration recently provides the best upper bounds~\cite{Sirunyan:2017xzt,Khachatryan:2016rke}
\begin{align}
  \mB(h\to e\mu)<3.5 \times 10^{-4}\,,
  \quad
  \mB(h \to e\tau)<6.1 \times 10^{-3}\,,
  \quad
  \mB(h \to \mu\tau)<2.5 \times 10^{-3}\,,
\end{align}
at 95\% CL, which have excluded the possibility of sizeable $\mu\tau$ flavor-violating Higgs interactions indicated by the previous CMS measurements~\cite{Khachatryan:2015kon}. 

The LFV Higgs couplings can also be indirectly constrained by the lepton FCNC processes, such as the $\mu \to e\gamma$ decay and $\mu \to e $ conversion in nuclei~\cite{He:2015rqa}. In the near future, the sensitivity for the branching ratio of $\mu \to e $ conversion in nuclei is expected to be improved by 4 orders of magnitude at the Mu2e experiment, i.e. from $7\times 10^{-13}$ in Au to $7 \times 10^{-17}$ in Al at 90\% CL~\cite{Abusalma:2018xem}.

It is also noted that, several hints of lepton-flavor university (LFU) violation emerge in the recent flavor physics data. The current experimental measurements on $R_{K^{(*)}}\equiv \mB(B \to K^{(*)} \mu^+ \mu^-)/\mB(B \to K^{(*)}e^+ e^-)$ and $R_{D^{(*)}}=\mB(B\to D^{(*)}\tau\nu)/\mB(B \to D^{(*)}\ell\nu)$ show about $2\sigma$~\cite{Aaij:2014ora,Aaij:2017vbb} and $4\sigma$~\cite{Amhis:2014hma} deviations from their SM predictions, respectively. Although such anomalies may not be related to the Higgs FCNC interactions directly, the NP candidates to explain these anomalies sometimes involve the Higgs FCNC couplings~\cite{Kim:2015zla,Chen:2013qta}.

In this work, motivated by these recent progresses and future prospects, we study the Higgs-mediated FCNC effects on various processes. We adopt an effective field theory (EFT) approach, in which the Higgs FCNC interactions are described by dim-6 operators~\cite{Harnik:2012pb}. In this approach, some FCNC couplings are severely constrained from flavor physics. In order to naturally obtain such small couplings, we concentrate on the minimal flavor violation hypothesis (MFV)~\cite{Chivukula:1987py,Buras:2000dm,DAmbrosio:2002vsn} as a particular working assumption. After deriving direct and indirect bounds on the Higgs FCNC couplings, we discuss in detail the future prospects of searching for these FCNC interactions in various processes.

This paper is organized as follows: In Sec.~\ref{sec:Higgs:Yukawa}, we give a brief overview of  the tree-level Higgs FCNC couplings in the EFT with and without the MFV hypothesis. In Sec.~\ref{sec:processes}, we discuss their effects on various flavor processes. In Sec.~\ref{sec:numerics}, we present our detailed numerical results and discussions. Our conclusions are given in Sec.~\ref{sec:conclusion}.

\section{Higgs FCNC Couplings}\label{sec:Higgs:Yukawa}
The Higgs FCNC Yukawa couplings appear in many extensions of the SM in the Higgs sector, such as multi-Higgs doublet models. In this work, we will not go into detailed model studies of these FCNC couplings but adopt an EFT approach to use known data to obtain model independent constraints on them. The framework that will be used for the analysis of  the Higgs FCNC couplings in the EFT approach and a special form in the MFV framework will be provided in the following.

\subsection{Higgs FCNC}\label{sec:Higgs:FCNC}

In the SM, the Yukawa interactions with quarks are described by the following Lagrangian in the interaction basis,
\begin{align}\label{eq:Lagrangian:SM}
  -\mathcal L_Y=\bar Q_L H Y_d d_R + \bar Q_L \tilde H Y_u u_R  +{\rm h.c.},
\end{align}
where $Q_L$ denotes the left-handed quark doublet, $d_R$ the right-handed down-type quarks, $u_R$ the right-handed up-type quarks, $H$ the Higgs doublet, and $\tilde H\equiv i\sigma_2 H^*$. The Yukawa coupling matrices $Y_{u,d}$ are $3\times 3$ complex matrices in flavor space. 

In the SM, the Higgs doublet develops a non-zero vacuum expectation value $\langle H \rangle = v /\sqrt{2}$ which breaks electroweak weak symmetry down to $U(1)_{\rm em}$, the charged Higgs fields $H^\pm$ and the imaginary part of the neutral components are ``eaten'' by $W^\pm$ and $Z$ bosons and left a physical neutral Higgs $h$. Working in the basis of quark mass eigenstates, the above Lagrangian gives a flavor conserving Higgs-fermion coupling of the form $m_f \bar f f (1+h/v)$.

When going beyond the SM, the above simple flavor conserving couplings will be modified. Considering the BSM effects in the EFT approach, these Higgs Yukawa interactions can be affected by dim-6 operators at the tree level. There are several different bases to choose for writing down the operators. We will work in the Warsaw basis in ref.~\cite{Grzadkowski:2010es}.   There exist only three operators relevant to our analysis to the lowest order. They are given by
\begin{align}\label{eq:operator}
  \mO_{dH} &= (H^\dagger H) (\bar Q_L H C_{dH} d_R) ,\nonumber\\
  \mO_{uH} &=(H^\dagger H) (\bar Q_L \tilde H C_{uH} u_R), \nonumber\\
  \mO_{\ell H} &=(H^\dagger H) (\bar L_L H C_{\ell H} e_R),
\end{align}
where the doublet/singlet $Q_L$, $u_R$, $d_R$, and the couplings $C_{uH,dH,\ell H}$ are in flavour space, and their flavour indices are omitted. There are other operators involving Higgs and fermions, such as operators involving $\bar H \overleftrightarrow{D} H$~\cite{Grzadkowski:2010es}. Such operators do not contribute FCNC at the tree level. Therefore we concentrate on the operators listed in the above. 

The operators above can contribute to the fermion mass terms in dim-4 $\mathcal L_\SM$ after the symmetry breaking $H^\dagger H \to 1/2 v^2$. The Yukawa couplings of $h$ to fermions are given by
\begin{align}
          \mathcal L_Y^f &= - \frac{1}{\sqrt{2}} \bar f_L \bar Y_f f_R v   - \frac{1}{\sqrt{2}} \bar f_L \biggl(\bar Y_f -\frac{v^2}{\Lambda^2} C_{fH} \biggr) f_R h +{\rm h.c.}\,.\nonumber\\
\end{align}
with the definition
\begin{align}\label{eq:Yuk:redef}
  \bar Y_f = Y_f - \frac{1}{2}\frac{v^2}{\Lambda^2}C_{fH},
\end{align}
where $\Lambda$ denotes some NP scale.

In the mass-eigenstate basis $\bar Y_f$ becomes diagonal, but the Higgs Yukawa interactions $(1/\sqrt{2}) (\bar Y_f -(v^2/\Lambda^2) C_{fH})$ is in general not diagonal~\cite{Chiang:2017etj} and induces FCNC interactions. We write them as
\begin{align}\label{eq:Lagrangian:FCNC}
  \mathcal L_Y = -\frac{1}{\sqrt{2}} \bar f (Y_L P_L + Y_R P_R) f h,
\end{align}
where $f$ denotes $(u,c,t)$, $(d,s,b)$ or $(e,\mu,\tau)$.  $Y_L$ and $Y_R$ are $3\times 3$ complex matrices in flavor space and connect to each other by the relation $Y_L = Y_R^\dagger$. In the SM, $Y_R^u =Y_f$ is diagonalized to have $\lambda_u^i = \sqrt{2}m_i / v$ and the vacuum expectation value $v = 246 \GeV$. Now $\bar Y_u$ plays the role of $Y_u$. Similarly for $d$ and $\ell$ sectors. Here we have used dim-6 operators to show how to parametrize the general form of a Higgs to fermions couplings. This should apply to more general cases.

In the literature, the following basis for the Higgs Yukawa interactions is also widely used
\begin{align}\label{eq:scalar basis}
  \mathcal L_Y = -\frac{1}{\sqrt{2}}  \bar f ( Y + i \gamma_5 \bar Y) f h\,.
\end{align}
Here, $Y$ and $\bar Y$ are $3\times 3$ Hermitian matrices. This form is related to eq.~\eqref{eq:Lagrangian:FCNC} by $Y_{R,L}=Y \pm i\bar Y $. It is noted that real $Y_{L,R}^{ij}$ do not imply real $Y_{ij}$ or $\bar Y_{ij}$, and \textit{vice versa}.

\subsection{Higgs FCNC in MFV}\label{sec:Higgs:FCNC:MFV}

In the SM, the Yukawa interactions in eq.~\eqref{eq:Lagrangian:SM} violate the global flavor symmetry
\begin{align}
  G_{\rm QF}=SU(3)_{Q_L}\otimes SU(3)_{u_R}\otimes SU(3)_{d_R}\,.
\end{align}
In the MFV~\cite{DAmbrosio:2002vsn} hypothesis, the flavor symmetry can be recovered by assuming the Yukawa couplings $Y_{u,d}$ to transform in the following representation
\begin{align}
  Y_u \sim (\boldsymbol{3}, \boldsymbol{\bar 3}, \boldsymbol{1})  \qquad \text{and} \qquad   Y_d \sim (\boldsymbol{3}, \boldsymbol{1}, \boldsymbol{\bar 3}).
\end{align}
Then, two basic building block spurions $\sfA\equiv Y_uY_u^\dagger$ and $\sfB\equiv Y_dY_d^\dagger$ under the group $SU(3)_{Q_L}\otimes SU(3)_{u_R}\otimes SU(3)_{d_R}$ transforming as $(\boldsymbol{1}+\boldsymbol{8},\boldsymbol{1},\boldsymbol{1})$ are important to parametrize the FCNC Yukawa couplings. Using polynomials of $\sfA$ and $\sfB$, which are denoted by $f(\sfA,\sfB)$, general forms of the $(\boldsymbol{3},\boldsymbol{\bar 3},\boldsymbol{1})$ and  $(\boldsymbol{3}, \boldsymbol{1}, \boldsymbol{\bar 3})$ tensors are $f_u(\sfA,\sfB)Y_u$ and $f_d(\sfA,\sfB)Y_d$, respectively. Therefore, to preserve the flavor symmetry $G_{\rm QF}$, the couplings in the effective operators of eq.~\eqref{eq:operator} should have the following forms
\begin{align}\label{eq:MFV:coupling}
  C_{dH}=f_d(\sfA,\sfB)Y_d \quad {\rm and} \quad C_{uH}=f_u(\sfA,\sfB)Y_u\,.
\end{align}
Using the Cayley-Hamilton identity, the polynomial $f(\sfA, \sfB)$ can be generally resumed into 17 terms~\cite{Colangelo:2008qp,Mercolli:2009ns},
\begin{align*}
  f(\sfA,\sfB)=\kappa_1 \mathds{1} +&\kappa_2 \sfA  + \kappa_5 \sfB^2 + \kappa_6 \sfA\sfB  + \kappa_8\sfA\sfB\sfA  +  \kappa_{11}\sfA \sfB^2 + \kappa_{13}\sfA^2\sfB^2 +\kappa_{15}\sfB^2\sfA\sfB+\kappa_{16}\sfA\sfB^2\sfA^2
  \\
  +& \kappa_3 \sfB+ \kappa_4 \sfA^2+ \kappa_7 \sfB\sfA +\kappa_{10} \sfB\sfA\sfB +\kappa_9 \sfB\sfA^2 + \kappa_{14}\sfB^2\sfA^2 + \kappa_{12}\sfA\sfB\sfA^2   +\kappa_{17}\sfB^2\sfA^2\sfB\,.
\end{align*}
Since the spurion $\sfB$ is highly suppressed by the small down-type quark Yukawa couplings, terms with $\sfB$ are neglected and we obtain~\cite{Chiang:2017hlj}
\begin{align}\label{eq:MFV:approx}
  f_u(\sfA,\sfB)  \approx \eu_0 \, \mathds{1} + \eu_1 \sfA + \eu_2 \sfA^2
  \quad{\rm and}\quad
  f_d(\sfA,\sfB) \approx \ed_0 \, \mathds{1} + \ed_1 \sfA + \ed_2 \sfA^2\,.
\end{align}
The coefficients $\eu_{0,1,2}$ and $\ed_{0,1,2}$ are free complex parameters but have negligible imaginary components~\cite{Colangelo:2008qp,Mercolli:2009ns,He:2014uya,He:2014efa,He:2014fva}.

For the down-type quarks, the Yukawa interactions with the dim-6 operator $\mO_{dH}$ after the EW symmetry breaking read
\begin{align}
          \mathcal L_Y^d &= - \frac{1}{\sqrt{2}} \bar d_L \bar Y_d d_R v   - \frac{1}{\sqrt{2}} \bar d_L \biggl(\bar Y_d -\frac{v^2}{\Lambda^2} C_{dH} \biggr) d_R h +{\rm h.c.}\,.\nonumber\\
\end{align}
with the definition
\begin{align}\label{eq:Yuk:redef}
  \bar Y_f = Y_f - \frac{1}{2}\frac{v^2}{\Lambda^2}C_{fH}\,.
\end{align}
Using the MFV hypothesis in eq.~\eqref{eq:MFV:coupling} and the approximation in eq.~\eqref{eq:MFV:approx},
\begin{align}
  C_{dH}= \bigl[ \ed_0 \,\mathds{1} + \ed_1 Y_u Y_u^\dagger + \ed_2 (Y_u Y_u^\dagger)^2 \bigr] Y_d\,.
\end{align}
With the redefinition in eq.~\eqref{eq:Yuk:redef}
\begin{align}
      C_{dH}= \bigl[ \ed_0 \,\mathds{1} + \ed_1 \bar Y_u \bar Y_u^\dagger + \ed_2 (\bar Y_u \bar Y_u^\dagger)^2 \bigr] \bar Y_d + \mO(v^2/\Lambda^2)\,.
\end{align}
Finally, we obtain the Yukawa interactions for down-type quarks in the mass eigenstate
\begin{align}
  \mathcal L_Y^d =  - \frac{1}{\sqrt{2}}   \bar d_L \bigl[ (1-\hed_0) \lambda_d - \hed_1 V^\dagger \lambda_u^2 V \lambda_d - \hed_2 V^\dagger \lambda_u^4 V \lambda_d \bigr]d_R h + {\rm h.c.}\,,
\end{align}
with the definition $\hed_i=(v^2/\Lambda^2)\ed_i$. Due to the large hierarchy in the diagonal matrix $\lambda_u$, the $\hed_1$ and $\hed_2$ terms have almost the same structure. Therefore, we will use the following approximation in the numerical analysis
\begin{align}\label{eq:Yuk:d}
      \mathcal L_Y^d = - \frac{1}{\sqrt{2}}   \bar d_L \bigl[ (1-\hed_0) \lambda_d - \hed_1 V^\dagger \lambda_u^2 V \lambda_d \bigr]d_R h + {\rm h.c.}\,,
\end{align}
which is equivalent to redefinite $(\hed_1 + \lambda_t^2 \hed_2) \to \hed_1$. We have checked that the numerical differences due to this approximation are negligible.

Similarly, the Yukawa interactions for up-type quarks in the MFV are obtained
\begin{align}
  \mathcal L_Y^u = - \frac{1}{\sqrt{2}}  \bar u_L \bigl[ (1-\heu_0)\lambda_u - \heu_1 \lambda_u^3 - \heu_2 \lambda_u^5 \bigr]u_R h + {\rm h.c.} \,,
\end{align}
with the definition $\heu_i = (v^2 /\Lambda^2)\eu_i$. Due to the large hierarchy in the diagonal matrix $\lambda_u$ and $\lambda_t\approx 1$, we take the approximation $\heu_0 \lambda_u + \heu_1 \lambda_u^3 + \heu_2 \lambda_u^5 \approx (\heu_0+\heu_1+\heu_2)\lambda_u$. Finally, after a redefinition $(\heu_0 + \heu_1 + \heu_2) \to \heu_0$, the following Lagrangian for up-type quarks are obtained
\begin{align}
  \mathcal L_Y^u =- \frac{1}{\sqrt{2}}  \bar u_L (1-\heu_0)\lambda_u u_R h + {\rm h.c.} \,.
\end{align}
We have checked that the numerical differences due to this approximation are negligible. In the MFV, the FCNC in the up sector is negligibly small.

For the lepton sector, definition of MFV depends on the underlying mechanism responsible for neutrino masses and is not unique~\cite{Cirigliano:2005ck,Cirigliano:2006su,Alonso:2011jd,Dinh:2017smk}. Here, we adopt the approach in ref.~\cite{Chiang:2017hlj}, which is based on type-I seesaw mechanism. Then, the basic building block spurion similar to $\sfA$ in the quark sector, reads in the mass eigenstate
\begin{align}\label{eq:A:ell}
  \sfA_\ell  = \frac{2\mathcal M}{v^2} U \hat m_\nu^{1/2}OO^\dagger \hat m_\nu^{1/2} U^\dagger\,,
\end{align}
where $U$ denotes the Pontecorvo-Maki-Nakagawa-Sakata matrix, $\hat m_\nu$ the diagonal neutrino mass matrix ${\rm diag}(m_1,m_2,m_3)$ and $\mathcal M$ mass of the right-handed neutrinos. Matrix $O$ is generally complex orthogonal, satisfying $O O^T = \mathds{1}$~\cite{Casas:2001sr}. Then, after neglecting small $\sfB_\ell$ terms, the Yukawa interactions for changed lepton reads
  \begin{align}
    \mathcal L_Y^\ell= - \frac{1}{\sqrt{2}} \bar \ell_L \left [(1-\hel_0)\lambda_\ell - \hel_1 \sfA_\ell \lambda_\ell - \hel_2 \sfA_\ell^2 \lambda_\ell \right ] \ell_R h\,,
  \end{align}
with the definition $\hel_i=(v^2/\Lambda^2)\el_i$.

In summary, the Yukawa couplings in the MFV framework can be written as in the basis of eq.~\eqref{eq:Lagrangian:FCNC},
\begin{align}
  Y_R^d &= (1-\hed_0)\lambda_d -  \hed_1  V^\dagger  \lambda_u^2  V \lambda_d \,,
  \nonumber\\
  Y_R^u &= (1-\heu_0)\lambda_u\,,
  \nonumber\\
  Y_R^\ell&=(1-\hel_0)\lambda_\ell - \hel_1 \sfA_\ell \lambda_\ell - \hel_2 \sfA_\ell^2 \lambda_\ell.
\end{align}
All the above Yukawa matrices are Hermitian in the MFV framework.

\section{Relevant Processes}\label{sec:processes}
In this section we consider possible processes which can constrain the Higgs FCNC couplings to fermions. We find the most relevant processes are $B_s-\bar B_s$, $B_d - \bar B_d$ and $K^0 -\bar K^0$ mixing, $B_{s,d}\to \ell_1 \ell_2$ decays, the leptonic decays $\ell_i \to \ell_j \gamma $ and $\mu \to e $ conversion in nuclei, and Higgs production and decay at the LHC, which are investigated in detail in this section.

\subsection{Neutral $\boldsymbol{B}$ and $\boldsymbol{K}$ meson mixing}

Including the Higgs FCNC contributions, the effective Hamiltonian for $B_s - \bar B_s$ mixing can be written as~\cite{Buras:2001ra}  
\begin{eqnarray}\label{eq:Heff:mixing}
\mathcal H_{\rm eff}^{\Delta B=2} 
= \frac{G_F^2}{16\pi^2}m_W^2 (V_{tb}V_{ts}^*)^2\sum_i  C_i \mathcal O_i + {\rm h.c.} ~,
\end{eqnarray}
where the operators relevant to our study are
\begin{align}
  \mathcal O_1^{\rm VLL}&=(\bar s^\alpha\gamma_\mu P_L b^\alpha)(\bar s ^\beta \gamma^\mu P_L b^\beta)\;,&
  \mathcal O_1^{\rm SLL}&=(\bar s^\alpha P_L b^\alpha)(\bar s^\beta P_L b^\beta)\;,\nonumber\\
  \mathcal O_2^{\rm LR}&=(\bar s^\alpha P_L b^\alpha)(\bar s^\beta P_R b^\beta)\;,&
  \mathcal O_1^{\rm SRR}&=(\bar s^\alpha P_R b^\alpha)(\bar s^\beta P_R b^\beta)\;,
\end{align}
with $\alpha$ and $\beta$ color indices. $V_{ij}$ denote the Cabibbo-Kobayashi-Maskawa (CKM) matrix elements. The SM contributes to only the $\mathcal O^{\rm VLL}_1$ operator, whose Wilson coefficients $C_1^{\rm VLL}$ can be found in ref.~\cite{Buchalla:1995vs}. The other operators can be generated by tree-level Higgs FCNC exchange, whose Wilson coefficients read~\cite{Chiang:2017etj}
\begin{align}\label{eq:WC:mixing}
    C_1^{\rm SLL, \NP} &= -\frac{1}{2}\tilde  \kappa \bigl(Y_L^{sb}\bigr)^2\;,&
    C_2^{\rm LR, \NP} &=-\tilde \kappa Y_L^{sb}Y_R^{sb}\;,
    \nonumber\\
    C_1^{\rm SRR, \NP} &= -\frac{1}{2}\tilde \kappa \bigl(Y_R^{sb}\bigr)^2\;,&
    \tilde \kappa  &=\frac{8\pi^2}{G_F^2} \frac{1}{m_h^2m_W^2}\frac{1}{(V_{tb}V_{ts}^*)^2}\;.
\end{align}

The contribution from $\mathcal H_{\rm eff}^{\Delta B=2}$ to the transition matrix element of $B_s - \bar B_s$ mixing is given by~\cite{Buras:2001ra},
\begin{align}
  M_{12}^s
= \langle B_s | \mathcal H_{\rm eff}^{\Delta B=2} | \bar B_s \rangle
= \frac{G_F^2}{16\pi^2} m_W^2 (V_{tb}V_{ts}^*)^2  \sum C_i \langle B_s \left\lvert \mathcal O_i \right\rvert \bar B_s \rangle\,,
\end{align}
where recent lattice calculations of the hadronic matrix elements $\langle \mathcal O_i \rangle$ can be found in refs.~\cite{Carrasco:2013zta,Bazavov:2016nty}. Then the mass difference and CP violation phase read
\begin{align}\label{eq:dltm}
  \Delta m_s = 2 |M_{12}^s|\,, \qquad \text{and} \qquad \phi_s= \arg M_{12}^s\,.
\end{align}
 In the case of complex Yukawa couplings, $\phi_s$ can derivate from the SM prediction, i.e., $\phi_s=\phi_s^\SM+\phi_s^\NP$. Nonzero $\phi_s^\NP$ can affect the CP violation in the $B_s \to J/\psi \phi$ decay~\cite{Artuso:2015swg}, as well as $\mathcal A_{\Delta\Gamma}$ in the $B_s \to \mu^+ \mu^-$ decay as in eq.~\eqref{eq:ys}. In the basis in eq.~\eqref{eq:scalar basis}, it can be seen that the mass difference $\Delta m_s$ depends only on $Y_{sb}^2$ and $\bar Y_{sb}^2$, but not $Y_{sb}\bar Y_{sb}$.  In addition, we follow ref.~\cite{Buras:2001ra} to perform renormalization group evolution of the NP operators $\mathcal O_1^{\rm SLL}$, $\mathcal O_1^{\rm SRR}$ and $\mathcal O_2^{\rm LR}$.  It is found that including RG effects of the NP operators enhances the NP contributions by about a factor of 2.

\subsection{$\boldsymbol{B_s \to \ell_1 \ell_2}$ decay}

In this subsection, we consider the $B_s \to \mu^+ \mu^-$ decay as an example to recapitulate the theoretical framework of the $B_s \to \ell_1 \ell_2$ processes. Within the Higgs FCNC effects, the effective Hamiltonian of the $B_s \to \mu^+ \mu^-$ decay reads~\cite{Buchalla:1995vs}
\begin{eqnarray}\label{eq:Hamiltonian}
  \mathcal H_{\rm eff}
  = -\frac{G_F}{\sqrt 2} \frac{\alpha_{e}}{\pi s_W^2} V_{tb}V_{ts}^*
  \bigl(C_A\mathcal O_A + C_S\mathcal O_S + C_P\mathcal O_P 
  + C'_S\mathcal O'_S + C'_P\mathcal O'_P \bigr)+{\rm h.c.},
\end{eqnarray}
where $\alpha_{e}$ is the fine structure constant, and $s_W^2 \equiv \sin^2\theta_W$ with $\theta_W$ being the weak mixing angle.  The operators $\mO_i^{(\prime)}$ are defined as
\begin{align}\label{eq:operator}
\mO_A &=\bigl(\bar q \gamma_\mu P_L b\bigr)\bigl(\bar\mu \gamma^\mu\gamma_5 \mu\bigr)\;,
&
\mO_S &= m_b \bigl(\bar q P_R b\bigr)\bigl(\bar \mu \mu\bigr)\;,
&
\mO_P &= m_b \bigl(\bar q P_R b\bigr)\bigl(\bar\mu \gamma_5 \mu\bigr)\;,\nonumber\\
&&
\mO_S^\prime&= m_b \bigl(\bar q P_L b\bigr)\bigl(\bar \mu \mu\bigr)\;,
&
\mO_P^\prime&= m_b \bigl(\bar q P_L b\bigr)\bigl(\bar\mu \gamma_5 \mu\bigr)\;.
\end{align}

In the framework we are working with, the Wilson coefficient $C_A$ contains only the SM contribution, and its explicit expression up to the NLO QCD corrections can be found in refs.~\cite{Buchalla:1993bv,Misiak:1999yg,Buchalla:1998ba}.  Recently, corrections at the NLO EW~\cite{Bobeth:2013tba} and NNLO QCD~\cite{Hermann:2013kca} have been completed, with the numerical value approximated by~\cite{Bobeth:2013uxa}
 \begin{eqnarray}
 &&C_A^{\SM}(\mu_b) = -0.4690 \left(\frac{m_t^{\rm P}}{173.1~\mbox{GeV}}\right)^{1.53} \left(\frac{\alpha_s(m_Z)}{0.1184}\right)^{-0.09}\;,
  \end{eqnarray}
where $m_t^{\rm P}$ denotes the top-quark pole mass.  In the SM, the Wilson coefficients $C_S^\SM$ and $C_P^\SM$ can be induced by the Higgs-penguin diagrams but are highly suppressed.  Their expressions can be found in refs.~\cite{Li:2014fea,Cheng:2015yfu}.  As a very good approximation, we can safely take $C_S^\SM=C_S^{\prime \SM}=C_P^\SM=C_P^{\prime \SM}=0$.

With the Higgs-mediated FCNC interactions in the effective Lagrangian, eq.~(\ref{eq:Lagrangian:FCNC}), the scalar and pseudoscalar Wilson coefficients 
\begin{align}
C_S^{\NP} &= \frac{1}{2}\kappa Y_R^{sb}  \bigl(Y_R^{\mu\mu}+Y_L^{\mu\mu}\bigr)\;,& C_P^{\NP} &=\frac{1}{2}\kappa Y_R^{sb} \bigl(Y_R^{\mu\mu}-Y_L^{\mu\mu}\bigr)\;, \nonumber\\
C_S^{\prime \NP} &= \frac{1}{2}\kappa Y_L^{sb} \bigl(Y_R^{\mu\mu}+Y_L^{\mu\mu}\bigr)\;,& C_P^{\prime \NP} &= \frac{1}{2}\kappa Y_L^{sb} \bigl(Y_R^{\mu\mu}-Y_L^{\mu\mu}\bigr)\;,
\end{align}
with the common factor 
\begin{align}
\kappa=\frac{\pi^2}{2G_F^2}\frac{1}{V_{tb}V_{ts}^*}\frac{1}{m_b m_h^2m_W^2}\;.
\end{align}

For the effective Hamiltonian eq.~\eqref{eq:Hamiltonian}, the branching ratio of $B_s \to \mu^+ \mu^-$ reads~\cite{Li:2014fea,Cheng:2015yfu}
\begin{eqnarray}
  \mathcal B(B_s \to \mu^+\mu^-) = 
  \frac{\tau_{B_s}G_F^4 m_W^4}{8\pi^5} |V_{tb}V_{ts}^*|^2 f_{B_s}^2 m_{B_s} m_\mu^2
  \sqrt{1-\frac{4 m_\mu^2}{m_{B_s}^2}} \bigl(|P|^2+|S|^2\bigr)\;,
\end{eqnarray}
where $m_{B_s}$, $\tau_{B_s}$ and $f_{B_s}$ denotes the mass, lifetime and decay constant of the $B_s$ meson, respectively. The amplitudes $P$ and $S$ are defined as
\begin{eqnarray}
  &&
  P\equiv
  C_A+\frac{m_{B_s}^2}{2 m_\mu}\left(\frac{m_b}{m_b+m_s}\right)(C_P-C_P^\prime)\;,\nonumber\\
  &&
  S \equiv 
  \sqrt{1-\frac{4m_\mu^2}{m_{B_s}^2}}\frac{m_{B_s}^2}{2 m_\mu}
  \left(\frac{m_b}{m_b+m_s}\right)(C_S-C_S^\prime)\;.
\end{eqnarray}
From these expressions and using the basis in eq.~\eqref{eq:scalar basis}, it can be seen that the branching ratio of $B_s \to \mu^+ \mu^-$ only depends on $\bar{Y}_{sb}Y_{\mu \mu}$ and $\bar{Y}_{sb}\bar{Y}_{\mu\mu}$.

Due to the $B_s$-$\bar B_s$ oscillations, the measured branching ratio of $B_s \to \mu^+ \mu^-$ should be the time-integrated one~\cite{DeBruyn:2012wk}:
\begin{align}
  \bar\mB(B_s \to \mu^+ \mu^-)=\left(\frac{1+\mathcal A_{\Delta\Gamma}y_s}{1-y_s^2}\right)\mathcal
  B(B_s\to \mu^+\mu^-)\;,
\end{align}
with~\cite{Buras:2013uqa}
\begin{align}\label{eq:ys}
  y_s = \frac{\Gamma_s^{\rm L}-\Gamma_s^{\rm H}}{\Gamma_s^{\rm L} + \Gamma_s^{\rm H}}=\frac{\Delta\Gamma_s}{2\Gamma_s}
  \quad \text{and} \quad
   \mathcal A_{\Delta\Gamma}=\frac{|P|^2\cos {(2\varphi_P-\phi_s^\NP)}-|S|^2\cos{( 2\varphi_S-\phi_s^\NP)}}{|P|^2+|S|^2}\;,
\end{align}
Here, $\Gamma_s^{\rm L}$ ($\Gamma_s^{\rm H}$) denote the decay widths of the light (heavy) $B_s$ mass eigenstates. $\varphi_P$ and $\varphi_S$ are the phases associated with $P$ and $S$, respectively. The CP phase $\phi_s^\NP$ comes from $B_s$-$\bar B_s$ mixing and has been defined in eq.~\eqref{eq:dltm}. In the SM, $\mathcal A_{\Delta\Gamma}^\SM=1$.

\subsection{Leptonic decays $\boldsymbol{\ell_i \to \ell_j \gamma}$}
Considering the Higgs FCNC interactions, the effective Lagrangian for the $\ell_i \to \ell_j \gamma$ decays are given by~\cite{Harnik:2012pb}
\begin{align}\label{eq:Leff:l_i_l_j_gm}
  \mathcal L_{\rm eff}=c_L \mathcal O_L + c_R \mathcal O_R + {\rm h.c.},
\end{align}
with the operators
\begin{align}
  \mathcal O_{L,R}=\frac{e}{8\pi^2}m_i (\bar\ell_j \sigma^{\mu\nu}P_{L,R}\ell_i)F_{\mu\nu}\,,
\end{align}
where $m_i$ denotes the mass of the lepton $\ell_i$ and $F_{\mu\nu}$ the photon field strength tensor. Then, the decay rate of $\ell_i \to \ell_j \gamma$ is given by~\cite{Harnik:2012pb}
\begin{align}
\Gamma(\ell_i \to \ell_j \gamma) = \frac{\alpha_e m_i^5}{64\pi^4}(|c_L|^2+|c_R|^2).
\end{align}
The Wilson coefficients $c_L$ and $c_R$ receive contributions from the one-loop penguin diagrams. Their analytical expressions read~\cite{Harnik:2012pb}
  \begin{align}
    c_L^{\rm 1-loop} = \sum_{f=e,\mu,\tau} F(m_i, m_f, m_j, 0, Y), 
    \qquad    
    c_R^{\rm 1-loop} = \sum_{f=e,\mu,\tau} F(m_i, m_f, m_j, 0, Y^\dagger),
  \end{align}
  with the loop function
  \begin{align*}
    F(m_i, m_f, m_j , q^2, Y)=\frac{1}{8m_i}\int_{0}^1 & dx dy dz \delta(1-x-y-z)
    \nonumber\\
    &\frac{xzm_jY_R^{jf}Y_L^{fi} + yzm_i Y_L^{jf}Y_R^{fi} + (x+y)m_f Y_L^{jf}Y_L^{fi}}    {zm_h^2-xzm_j^2-yzm_i^2 +(x+y)m_f^2 - xy q^2}.
  \end{align*}
At the two-loop level, there are also comparable contributions from the Barr-Zee type diagrams. Here, we use the numerical results in ref.~\cite{Harnik:2012pb}.
  \begin{align}\label{eq:two-loop}
    c_L^{\rm 2-loop} &\approx \frac{1}{\sqrt 2 m_h^2}\frac{m_\tau}{m_i} Y_L^{ji} (-0.058 Y_R^{tt}+0.11),
    \nonumber\\
    c_R^{\rm 2-loop} &\approx \frac{1}{\sqrt 2 m_h^2}\frac{m_\tau}{m_i} Y_R^{ji} (-0.058 Y_L^{tt}+0.11),
  \end{align}
which are obtained from a full two-loop analytical calculations~\cite{Chang:1993kw}. Here, $Y_L^{tt}$ and $Y_R^{tt}$ are assumed to be real.


\subsection{$\boldsymbol{\mu\to e}$ conversion in nuclei}
The Higgs FCNC interactions could induce $\mu \to e$ conversion when $\mu$ is in nuclei. The relevant effective Lagrangian reads~\cite{Harnik:2012pb}
\begin{align}
  \mathcal L_{\rm eff}  = c_L \frac{e}{8\pi^2}m_\mu (\bar e \sigma^{\mu \nu} P_L \mu) F_{\mu\nu}
  -\frac{1}{2}\sum_q \Bigl [g_{LS}^q(\bar e P_R \mu)(\bar q q) + g_{LV}^q(\bar e \gamma^\mu P_L \mu)(\bar q \gamma_\mu q)  \Bigr]+ \bigl(L\leftrightarrow R\bigr),
\end{align}
where the summation runs over all quark flavors $q\in \lbrace u,d,s,c,b,t\rbrace$. The Wilson coefficient $c_{L,R}$ are the same with the ones in $\mu \to e\gamma$ in eq.~\eqref{eq:Leff:l_i_l_j_gm}. The scalar operators are generated by the tree-level Higgs exchange and their Wilson coefficients are given by
  \begin{align}
    g_{LS}^q = -\frac{1}{m_h^2}Y_R^{e\mu}{\rm Re}(Y_R^{qq}), \qquad g_{RS}^q = -\frac{1}{m_h^2}Y_{L}^{e \mu} {\rm Re}(Y_{R}^{qq}).
  \end{align}
For the vector operators, the leading contributions arise from one-loop penguin diagrams, whose Wilson coefficients read~\cite{Harnik:2012pb}
\begin{align}\label{eq:WC:scalar}
    g_{LV}^q=-\frac{\alpha_e Q_q}{2\pi q^2}\sum_{f=e,\mu,\tau}[G(m_\mu, m_f, m_e,q^2,Y)-G(m_\mu, m_f, m_e, 0,Y)],
\end{align}
with the loop function
\begin{align}
  G(m_i, m_f, m_j, q^2, Y) =\frac{1}{2} \int_0^1 &dx \int_0^{1-x} dy \biggl\lbrace+Y_R^{jf}Y_L^{fi}\log\Delta - \frac{1}{\Delta}\bigl(m_i m_j z^2 Y_L^{jf}Y_R^{fi}\bigr) 
\\
-&\frac{1}{\Delta}\Bigl[m_f m_j z Y_L^{jf}Y_L^{fi} +m_fm_iz Y_R^{jf}Y_R^{fi} +\bigl(q^2 xy +m _f^2\bigr) Y_R^{jf} Y_L^{fi} \Bigr]\biggr\rbrace, \nonumber
\end{align}
where $\Delta \equiv z m_h^2 - xz m_j^2 -yz m_i^2 +(x+y)m_f^2 - xyq^2$ and $z\equiv 1-x-y$. Here, $Q_q$ is the charge of quark $q$. $q^2$ denotes square of the moment exchange and takes the value of $- m_\mu^2$, which corresponds to the limit of an infinitely heavy nucleus. The coupling $g_{RV}^q$ can be obtained from $g_{LV}^q$ with the replacement $Y \to Y^\dagger$.

Using these Wilson coefficients, the rate of $\mu \to e$ conversion in a nuclei $N$ can be written as~\cite{Kitano:2002mt}
\begin{align}
  \Gamma(\mu N \to e N)   = \left\lvert - \frac{e}{16\pi^2} c_R D + \tilde g_{LS}^{(p)}S^{(p)} + \tilde g_{LS}^{(n)} S^{(n)} + \tilde g_{LV}^{(p)} V^{(p)} \right\rvert^2 + \bigl(L \leftrightarrow R \bigr).
\end{align}
Here, $\tilde g_{L/RS,L/RV}^{(n,p)}$ denote the couplings to proton and neutron and can be evaluated from the quark-level ones
  \begin{align}
        \tilde g_{LS,RS}^{(p)}=&\sum_q g_{LS,RS}^q\frac{m_p}{m_q}f^{(q,p)}\,,
        \quad\quad
        \tilde g_{LV,RV}^{(p)} = g_{LV,RV}^q/Q_q\,,
        \nonumber\\
        \tilde g_{LS,RS}^{(n)}=&\sum_q g_{LS,RS}^q\frac{m_n}{m_q}f^{(q,n)}\,,
  \end{align}
where the summation runs over all quark flavors $q\in \lbrace u,d,s,c,b,t \rbrace$, and the nucleon matrix elements $f^{(q,p)}\equiv \langle p \lvert m_q \bar q q \rvert p \rangle / m_p$ are numerically~\cite{Ellis:2008hf,Young:2009ps}
\begin{align}
  f^{(u,p)}=f^{(d,n)}&=0.024, &f^{(c,p)}=f^{(b,p)}=f^{(t,p)}=\frac{2}{27}\Bigl( 1-\sum_{q=u,d,s}f^{(q,p)}\Bigr),
  \nonumber\\
  f^{(d,p)}=f^{(u,n)}&=0.033, &f^{(c,n)}=f^{(b,n)}=f^{(t,n)}=\frac{2}{27}\Bigl( 1-\sum_{q=u,d,s}f^{(q,n)}\Bigr),
  \nonumber\\
  f^{(s,p)}=f^{(s,n)}&=0.25.
\end{align}
The coefficients $D$, $V^{(p)}$, $S^{(p)}$, and $S^{(n)}$ denote overlap integrals of the muon, electron and nuclear wave function. For the Au and Al nuclei, their values read~\cite{Kitano:2002mt}
\begin{align}
 \bigl(D,V^{(p)},S^{(p)},S^{(n)}\bigr)=
 \begin{cases}
   0.1890, 0.0974, 0.0614, 0.0918, & {\rm for\;Au,}\\
   0.0362, 0.0161, 0.0155, 0.0167, & {\rm for\;Al,}
  \end{cases}
\end{align}
in unit of $m_\mu^{5/2}$.

Finally, the branching ratio of $\mu \to e$ conversion are obtained 
\begin{align}
\mathcal B( \mu N \to eN) = \frac{\Gamma(\mu N \to e N)}{\Gamma_{{\rm capt.}\,N}},
\end{align}
where $\Gamma_{{\rm capt.}\,N}$ denotes the muon capture rate, and numerically $\Gamma_{\rm capt.\,Au} = 1.307 \times 10^7 {\, \rm s}^{-1}$ and $\Gamma_{\rm capt.\,Al}=7.054\times 10^5 {\,\rm s}^{-1}$~\cite{Suzuki:1987jf}.

\section{Numerical Analysis}\label{sec:numerics}
In this section, we proceed to present our numerical analysis for the Higgs FCNC couplings in the general case and in the MFV framework in Sec.~\ref{sec:Higgs:FCNC} and Sec.~\ref{sec:Higgs:FCNC:MFV}, respectively. Tab.~\ref{tab:input} shows the relevant input parameters, and Tab.~\ref{tab:exp} summarises the SM predictions and the current experimental data for various processes discussed in the previous sections.

To constrain the Higgs FCNC couplings, we impose the experimental constraints in the same way as in ref.~\cite{Jung:2012vu,Chiang:2017etj}; i.e., for each point in the parameter space, if the difference between the corresponding theoretical prediction and experimental data is less than $1.96\,\sigma$ ($1.65\,\sigma$) error bar, which is calculated by adding the theoretical and experimental errors in quadrature, this point is regarded as allowed at 95\% CL (90\% CL). Since the main theoretical uncertainties arise from hadronic input parameters, which are common to both the SM and the Higgs FCNC contributions, the relative theoretical uncertainty is assumed to be constant over the whole parameter space.

\begin{table}[t]
  \centering
  \begin{tabular}{lllll l}
    \hline\hline
    Input & Value & Unit & Ref. 
  \\\hline
  $\alpha_s^{(5)}(m_Z)$ & $0.1181\pm 0.0011$ & & \cite{PDG:2018}
  \\
  $m_t^{\rm P}$ & $173.1 \pm 0.9$ & GeV & \cite{PDG:2018}
  \\
  \hline
  $|V_{cb}|$ (semi-leptonic) & $41.00 \pm 0.33 \pm 0.74$ &$10^{-3}$& \cite{Charles:2004jd}
  \\
  $|V_{ub}|$ (semi-leptonic) & $3.98 \pm 0.08 \pm 0.22$ & $10^{-3}$& \cite{Charles:2004jd}
  \\
  $|V_{us}|f_+^{K\to\pi}(0)$ &  $0.2165 \pm 0.0004$ & & \cite{Charles:2004jd}
  \\
  $\gamma$ &  $72.1_{-5.8}^{+5.4}$ & $[^\circ]$ & \cite{Charles:2004jd}
  \\
  $f_+^{K\to \pi}(0)$ & $0.9681 \pm 0.0014 \pm 0.0022$ & & \cite{Charles:2004jd}
  \\
  \hline
  $\sin^2\theta_{12}$ & $0.307_{-0.012}^{+0.013}$ & & \cite{Esteban:2016qun}
  \\
  $\sin^2\theta_{23}$ & $0.538_{-0.069}^{+0.033}$ ($0.554_{-0.033}^{+0.023}$) & & \cite{Esteban:2016qun}
  \\
  $\sin^2\theta_{13}$ & $0.02206_{-0.00075}^{+0.00075} $ ($0.02227_{-0.00074}^{+0.00074}$) & & \cite{Esteban:2016qun}
  \\
  $\delta_{\rm CP}$ & $234_{-31}^{+43}$ ($278_{-29}^{+26}$) & $[^\circ]$ & \cite{Esteban:2016qun}
  \\
  $\Delta m_{21}^2$ & $7.40_{-0.20}^{+0.21}$ & $10^{-5}{\,\rm eV}^2$ & \cite{Esteban:2016qun}
  \\
  $\Delta m_{3\ell}^2$ & $+2.494_{-0.031}^{+0.033}$ ($-2.465_{-0.031}^{+0.032}$)& $10^{-3}{\,\rm eV}^2$ & \cite{Esteban:2016qun}
  \\
  \hline
  $f_{B_s}$ & $228.4\pm 3.7$ & MeV & \cite{Aoki:2016frl}
  \\
  $f_{B_d}$ & $192.0\pm 4.3$ & MeV & \cite{Aoki:2016frl}
  \\
  $f_K$     & $155.7\pm 0.7$  & MeV & \cite{Aoki:2016frl}
  \\
  $f_{B_s}\sqrt{\hat B_s}$ & $274\pm 8$ & MeV & \cite{Aoki:2016frl}
  \\
  $f_{B_d}\sqrt{\hat B_d}$ & $225\pm 9$ & MeV & \cite{Aoki:2016frl}
  \\
  $\hat B_K$ & $0.7625 \pm 0.0097$ & & \cite{Aoki:2016frl}
  \\
  $1/\Gamma_s^{\rm H}$  & $1.609 \pm 0.010$ & ps & \cite{Amhis:2016xyh}
  \\
  $\Delta\Gamma_s/\Gamma_s$ & $0.128\pm 0.009$ & & \cite{Amhis:2016xyh}
  \\
  \hline\hline
  \end{tabular}
  \caption{Input parameters used in the numerical analysis. The neutrino oscillation parameters (values in brackets) correspond to the normal (inverted) Ordering of the light neutrinos' masses.}
  \label{tab:input}
\end{table}

\begin{table}[t]
  \centering
  \begin{tabular}{l l l l l}
    \hline\hline
    OBSERVABLE & SM & EXP & Ref \\
    \hline
    $\mB(h \to e \mu)$ & ~~- & $<3.5 \times 10^{-4}$ & \cite{Khachatryan:2016rke}
    \\
    $\mB(h \to e \tau)$ & ~~- & $<6.1 \times 10^{-3}$ & \cite{Sirunyan:2017xzt}
    \\
    $\mB(h \to \mu\tau)$ & ~~- & $<2.5 \times 10^{-3}$ & \cite{Sirunyan:2017xzt}
    \\
    \hline
    $\mB(\mu\to e\gamma)\vphantom{\frac{1}{2}_|^|}$ &~~-& $<4.2\times10^{-13}$ &\cite{TheMEG:2016wtm}
    \\
    $\mB(\tau\to e\gamma)\vphantom{\frac{1}{2}_|^|}$ &~~-& $<3.3\times10^{-8}$ &\cite{PDG:2018}
    \\
    $\mB(\tau\to\mu\gamma)\vphantom{\frac{1}{2}_|^|}$ &~~-& $<4.4\times10^{-8}$ &\cite{PDG:2018}
    \\ \hline
   $\mB(\mu\to eee)\vphantom{\frac{1}{2}_|^|}$&~~-& $<1.0\times10^{-12}$ &\cite{PDG:2018}
   \\
    $\mB(\tau\to eee)\vphantom{\frac{1}{2}_|^|}$ &~~-&  $<2.7\times10^{-8}$& \cite{PDG:2018}
    \\
    $\mB(\tau\to\mu\mu\mu)\vphantom{\frac{1}{2}_|^|}$ &~~-& $<2.1\times10^{-8}$ &\cite{PDG:2018}
    \\ \hline
    $\mB(\mu{\rm Au}\to e{\rm Au})\vphantom{\frac{1}{2}_|^|}$ &~~-& $<7.0\times10^{-13}$ &\cite{Bertl:2006up}
    \\\hline
    $\overline{\mathcal B}(B_s \to \mu^+ \mu^-)[\,10^{-9}]$& $3.43 \pm 0.19$ & $3.1 \pm 0.7$ & \cite{Amhis:2016xyh}
    \\
    $\Delta m_d [\,{\rm ps}^{-1}]$ &  $0.607_{-0.075}^{+0.075}$ & ~\, $0.5064 \pm 0.0019 $ & \cite{Amhis:2016xyh}
    \\
    $\Delta m_s [\,{\rm ps}^{-1}]$ &  $19.196_{-1.341}^{+1.377}$ & $17.757 \pm 0.021 $ & \cite{Amhis:2016xyh}
    \\
    $\phi_s [\,{\rm rad}]$ & $-0.042_{-0.003}^{+0.003}$ & $-0.021 \pm 0.031$ & \cite{Amhis:2016xyh}
    \\
    $\Delta m_K [\,10^{-3}\,{\rm ps}^{-1}]$ & ~\,$4.68\pm 1.88$ &~\,$5.293 \pm 0.009$& \cite{PDG:2018}
    \\
    $\lvert \varepsilon_K\rvert [10^{-3}]$ & ~\,$2.33_{-0.29}^{+0.27}$ & ~\,$2.228\pm 0.011$& \cite{PDG:2018}
    \\\hline\hline
  \end{tabular}
  \caption{\baselineskip 3.0ex
  SM predictions and experimental measurements for the observables used in the numerical analysis. Upper bounds for the Higgs LFV decays are values corresponding to 95\%\,CL, while the other LFV processes 90\%\,CL.}
  \label{tab:exp}
\end{table}

\subsection{Analysis within general Higgs FCNC}
In our previous paper~\cite{Chiang:2017etj}, the Higgs FCNC interactions in eq.~\eqref{eq:Lagrangian:FCNC} have already been studied in detail. Here, we focus on the couplings $Y_{L,R}^{e \mu}$ and $Y_{L,R}^{e\tau}$, which have not been investigated previously. These two couplings could induce $h\to e \mu$ and $h \to e \tau$ decay, respectively. The current Higgs data give the following bounds
\begin{align}
  \sqrt{\lvert Y_L^{e\mu} \rvert^2 + \lvert Y_R^{e \mu} \rvert^2}< 7.2 \times 10^{-4}\,,
  \qquad {\rm and} \qquad
  \sqrt{\lvert Y_L^{e\tau} \rvert^2 + \lvert Y_R^{e \tau} \rvert^2}< 3.0 \times 10^{-3}\,, 
\end{align}
at 95\% CL. When obtaining these bounds, the contributions of $Y_{L,R}^{e\mu}$ and $Y_{L,R}^{e\tau}$ to the Higgs total width have been included.

\begin{figure}[t]
  \centering
  \includegraphics[width=0.45\textwidth]{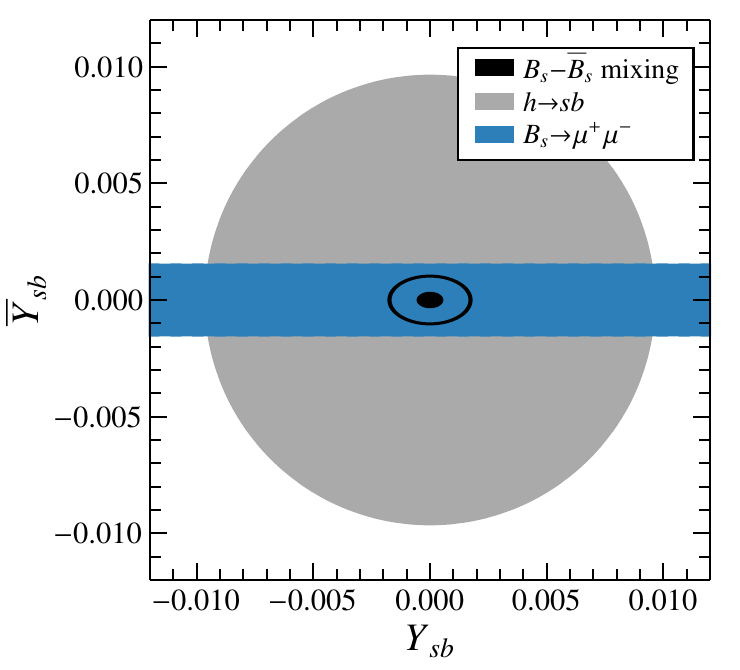}
  \caption{Allowed region of $(Y_{sb},\,\bar Y_{sb})$ at 95\%\,CL, assuming real $Y_{sb}$ and $\bar Y_{sb}$ couplings. The black region corresponds to the allowed parameter space by $B_s - \bar B_s$ mixing. The blue region is allowed by $\mB(B_s \to \mu^+ \mu^-)$ with the assumption $(Y_{\mu\mu},\, \bar Y_{\mu\mu})=(Y_{\mu\mu}^\SM,\, 0)$. In the dark region, $\Gamma(h\to sb)<1.4\MeV$.}
  \label{fig:Bs mixing}
\end{figure}

The FCNC couplings $Y_{sb}$ and $\bar Y_{sb}$ are constrained by $B_s - \bar B_s$ mixing. In the case of real $Y_{sb}$ and $\bar Y_{sb}$, their allowed regions by $\Delta m_s$ are shown in Fig.~\ref{fig:Bs mixing}. There are two allowed regions. The one near the origin corresponds to the case where the Higgs FCNC effects are destructive with the SM contribution. In the other region, the Higgs-mediated FCNC interactions dominate over the SM contribution. Another bound on these two parameters comes from the $h \to sb$ decay. Although there is no upper limits on this process currently, we consider the bound $\mB(h \to {\rm new})<34\%$ at 95\%CL obtained at the LHC Run I~\cite{Khachatryan:2016vau}, which denotes the upper limit on the overall branching fraction of the Higgs boson into BSM decays. However, this constraint is much weaker than the one from $B_s -\bar B_s$ mixing, as shown in Fig.~\ref{fig:Bs mixing}. Furthermore, assuming a SM-like $h\mu\mu$ coupling, $\mB(B_s \to \mu^+ \mu^-)$ also provides a constraint on $\bar Y_{sb}$. Such constraints is comparable with the one from $B_s - \bar B_s$ mixing, as can be seen in Fig.~\ref{fig:Bs mixing}. In the case of complex $Y_{sb}$ and $\bar Y_{sb}$, situation becomes quite different. Since the contributions of $Y_{sb}$ and $\bar Y_{sb}$ to $\Delta m_s$ can cancel to each other, $B_s - \bar B_s$ mixing can't provide upper limits on $|Y_{sb}|$ and $|\bar Y_{sb}|$. In this case, the upper bounds are given by $\mB(B_s \to \mu^+ \mu^-)$ with the assumption of a SM-like $h\mu\mu$ coupling and are weaker than the ones in the case of real couplings. Finally, the combined constraints on the complex couplings $Y_{sb}$ and $\bar Y_{sb}$ result in the following prediction
\begin{align*}
  \Gamma(h \to s b) < 0.17 \MeV,
\end{align*}
at 95\% CL.

For the $B_s \to \ell_1 \ell_2$ decays, using the analytical expressions in Sec.~\ref{sec:processes}, we can obtain the following numerical expression
\begin{align}
  \frac{\mB(B_s \to \ell_1 \ell_2)}{\mB(h \to \ell_1 \ell_2)} \approx 2.1|\bar Y_{sb}|^2\,,
\end{align}
where the SM Higgs total width $\Gamma_h^{\rm SM} \approx 4.07\MeV$~\cite{Heinemeyer:2013tqa} is assumed. In the case of complex Yukawa couplings, the combined bounds on $Y_{sb}$ and $\bar Y_{sb}$ discussed above and the LHC bounds on $h \to \ell_i \ell_j$ result in following upper limits
\begin{align*}
  \mB(B_s \to e \mu) < 2.1 \times 10^{-9}\,,
  \quad
  \mB(B_s \to e \tau) < 3.7 \times 10^{-8}\,,
  \quad
  \mB(B_s \to \mu \tau) < 1.5 \times 10^{-8}\,,
\end{align*}
at 95\% CL. For the branching ratio of $B_s \to e \mu$ decay, our predicted upper limit is three times lower than the current LHCb bound $\mB(B_s \to e \mu)<6.3 \times 10^{-9}$~\cite{Aaij:2017cza}.

The Higgs FCNC couplings can also affect the LFV processes in the lepton sector, such as the $\mu \to e \gamma$ decay. However, their dominated contributions arise at loop level and involve several Yukawa couplings. These processes can't provide model-independent bounds on one or two particular Yukawa couplings except assuming some special hierarchy among the Higgs FCNC couplings $Y_{L,R}^{i,j}$, as in ref.~\cite{Harnik:2012pb}.

\subsection{Analysis in the MFV framework}
The Higgs FCNC couplings in the MFV framework have been discussed in detail in Sec.~\ref{sec:Higgs:FCNC:MFV}. In the following numerical analysis, without loss of generality, we take the NP scale $\Lambda=v$, such that $\hat{\epsilon}_{0,1,2}^{u,d,\ell}=\epsilon_{0,1,2}^{u,d,\ell}$, and the right-handed neutrinos' mass $\mathcal M=10^{15}\GeV$. For the MFV in the lepton sector, we consider the simplest possibility that the orthogonal matrix $O$ in eq.~\eqref{eq:A:ell} is real. Since mass ordering of light neutrinos is not yet established, both the normal ordering (NO), where $m_1 < m_2 < m_3$, and the inverted ordering (IO), where $m_3 < m_1 < m_2$, are included in our analysis. In the NO (IO) case, we take $m_{1(3)}=0$. Finally, the Higgs Yukawa couplings in the MFV framework are determined by the following 6 real parameters
\begin{align}
\bigl( \eu_0, \; \ed_0, \; \ed_1, \;  \el_0, \; \el_1, \; \el_2     \bigr),
\end{align}
which correspond to the up-type quark, down-type quark and lepton sectors, respectively. In the following, the constraints on these parameters will be discussed in detail.

\begin{figure}[t]
  \centering
  \includegraphics[width=0.4\textwidth]{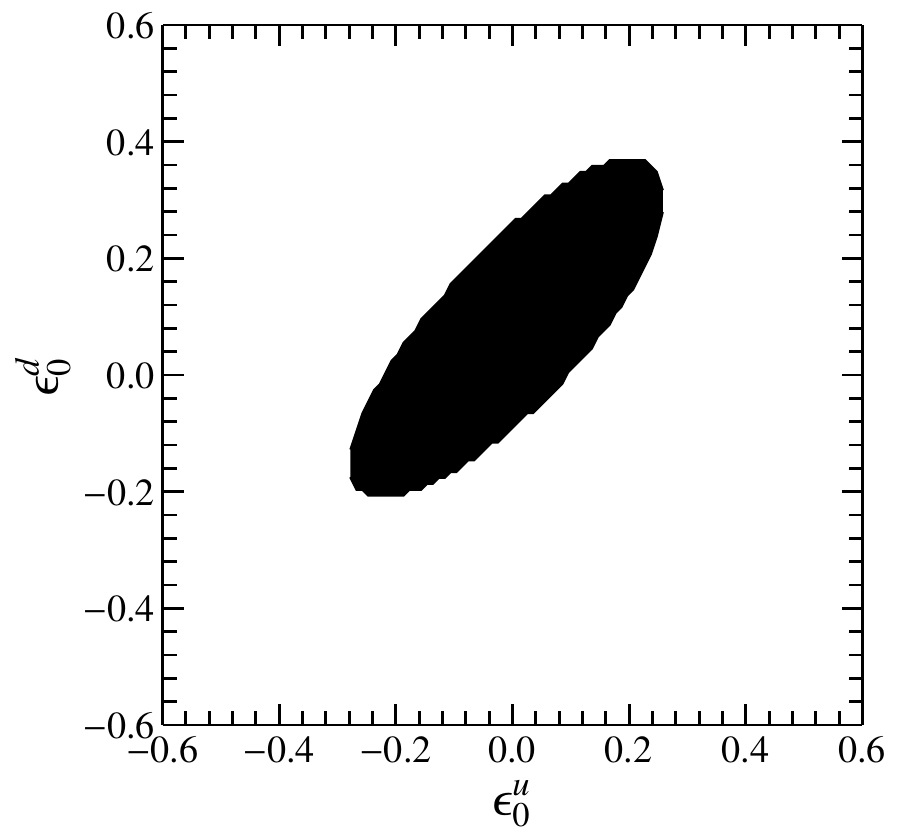}
  \qquad
  \includegraphics[width=0.4\textwidth]{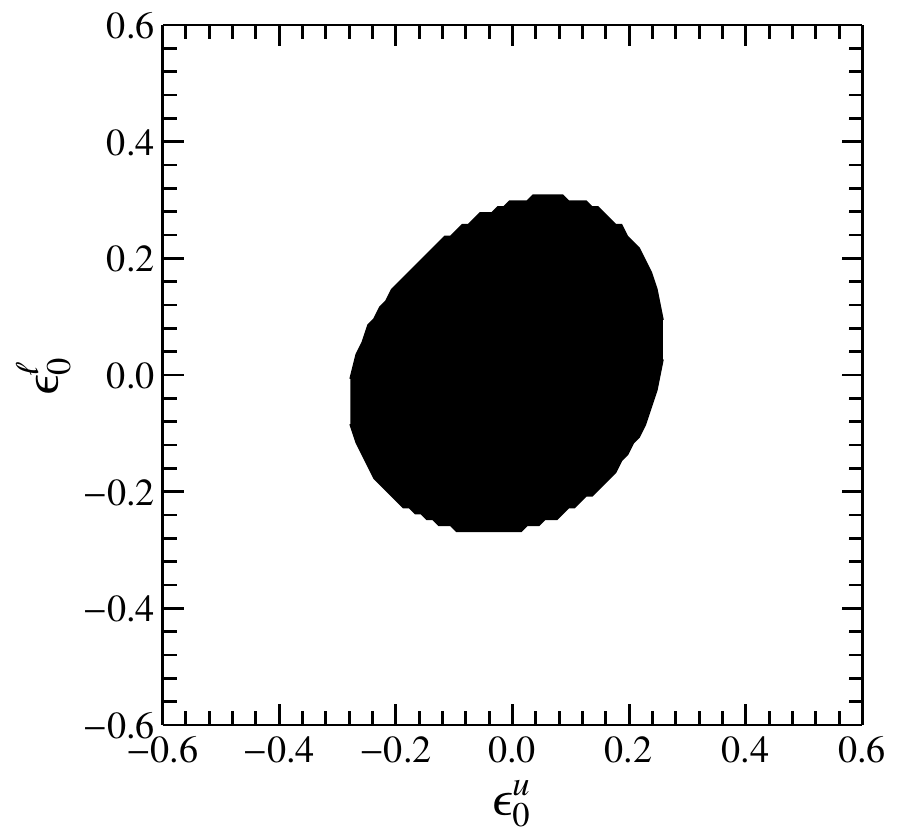}
  \caption{Allowed region of $(\eu_0, \, \ed_0, \,\el_0) $ by the LHC Higgs data at 90\% CL, plotted in the $(\eu_0,\,\ed_0)$ (left) and $(\eu_0, \, \el_0)$ (right) plane.}
  \label{fig:eu0:ed0:el0}
\end{figure}

The parameters $(\eu_0, \, \ed_0,  \, \el_0)$ control the Higgs flavor-conserving couplings to up-type quarks, down-type quarks and leptons, respectively. They are constrained by the Higgs production and decays processes at the LHC. We perform a global fit for these three parameters with the \texttt{Lilith} package~\cite{Bernon:2015hsa}, which is used to take into account the Higgs data measured by LHC Run I~\cite{Khachatryan:2016vau} and Tevatron~\cite{Aaltonen:2013ioz}. Although the flavor-changing parameters $\ed_1$ and $\el_{1,2}$ can also affect the Higgs signal strengths, they are strongly bounded by other processes, as discussed in the following. Therefore, their contributions can be safely neglected in the global fit. The allowed regions of $(\eu_0,\, \ed_0, \, \el_0)$ at 90\% CL are shown in Fig.~\ref{fig:eu0:ed0:el0}. Our global fit shows that $\mO(30\%)$ deviations from the SM values are allowed for the flavor-conserving couplings in the MFV framework.

The flavor-changing couplings for down-type quarks are determined by the parameter $\ed_1$. Constraints on this coupling come from $B_s - \bar B_s$, $B_d - \bar B_d$ and $K^0 - \bar K^0$ mixing. Since hadronic uncertainties in $K^0 -\bar K^0$ mixing are relatively large~\cite{Buras:2013ooa,Bertolini:2014sua}, we adopt the conservative treatment in ref.~\cite{Bertolini:2014sua}; i.e., the Higgs FCNC effects to $\Delta m_K$ are allowed within 50\% range of $\Delta m_K^{\rm exp}$, and $|\epsilon_K|$ is allowed to vary within a 20\% symmetric range. Since the current experimental data of the $B$ and $K$ mixing are in good agreement with the SM prediction, we obtain the strong bound on the MFV parameter 
\begin{align}
  |\ed_1 |<0.59,
\end{align}
at 95\% CL. This bound is dominated by $\Delta m_s$ in $B_s - \bar B_s$ mixing. Since the Yukawa couplings $Y_{L,R}^{sd}$ in the MFV framework are suppressed by $s$ or $d$ quark mass as in eq.~\eqref{eq:Yuk:d}, $K^0- \bar K^0$ mixing can't provide strong constraint. Using this bound, the predicted upper limits for various Higgs FCNC decays are obtained
\begin{align}
  \Gamma(h \to sd)&< 7.4 \times 10^{-11}\MeV,
  \nonumber\\
  \Gamma(h \to sb)&< 2.0 \times 10^{-3}\:\,\MeV,
  \nonumber\\
  \Gamma(h \to db)&< 9.4 \times 10^{-5}\:\,\MeV,
\end{align}
at 95\% CL. Such small decay rates make these channels very difficult to measure at the LHC~\cite{Barducci:2017ioq}.

The parameters $(\el_1, \, \el_2)$ control the flavor-changing couplings for changed leptons. They should be bounded by the LFV processes. However, as discussed in Sec.~\ref{sec:processes}, the quark Yukawa couplings also appear in some leptonic processes, e.g., top quark Yukawa couplings are involved in the two-loop diagrams of $\mu \to e \gamma$ and all the quark Yukawa couplings affect $\mu\to e $ conversion in nuclei at the tree level. Generally, all relevant parameters in the LFV processes are $(\eu_0,\,\ed_0,\,\el_0,\, \el_1, \,\el_2)$. We don't include the MFV parameter $\ed_1$, since its effect is highly suppressed in the LFV processes. When deriving the bounds on these parameters and studying their effects, it's useful to separate from the effects of the quark Yukawa couplings. Therefore, we consider the following two scenarios in the discussion of the LFV processes.
\begin{align}
  {\rm Scenario \; I:} & -0.5 < \el_{0,1,2} <+0.5, & {\rm Scenario \; II:} & -1.0 < \el_{0,1,2}<+1.0, 
  \nonumber\\
  & \qquad\eu_0=\ed_0=0, & & -1.0 < \, \epsilon_0^{u,d} \; < +1.0, 
\end{align}
Scenario I corresponds to the case that the flavor-conserving quark Yukawa couplings are the SM-like, and Scenario II the most general case.

\begin{figure}[t]
  \centering
    \includegraphics[width=0.4\textwidth]{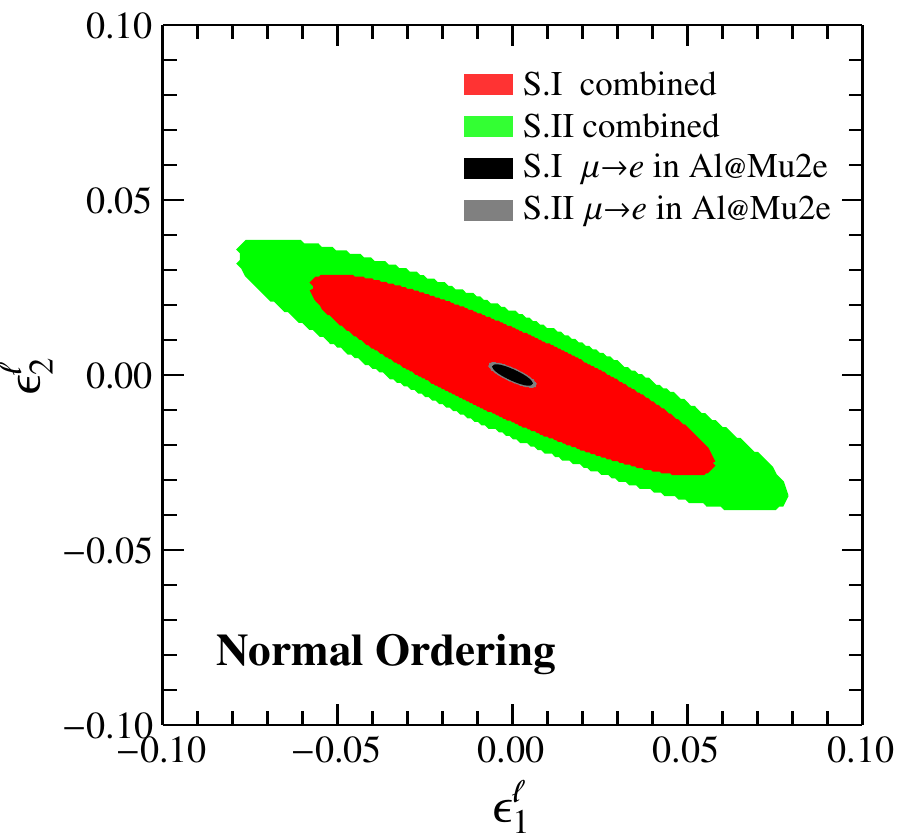}
    \qquad
    \includegraphics[width=0.4\textwidth]{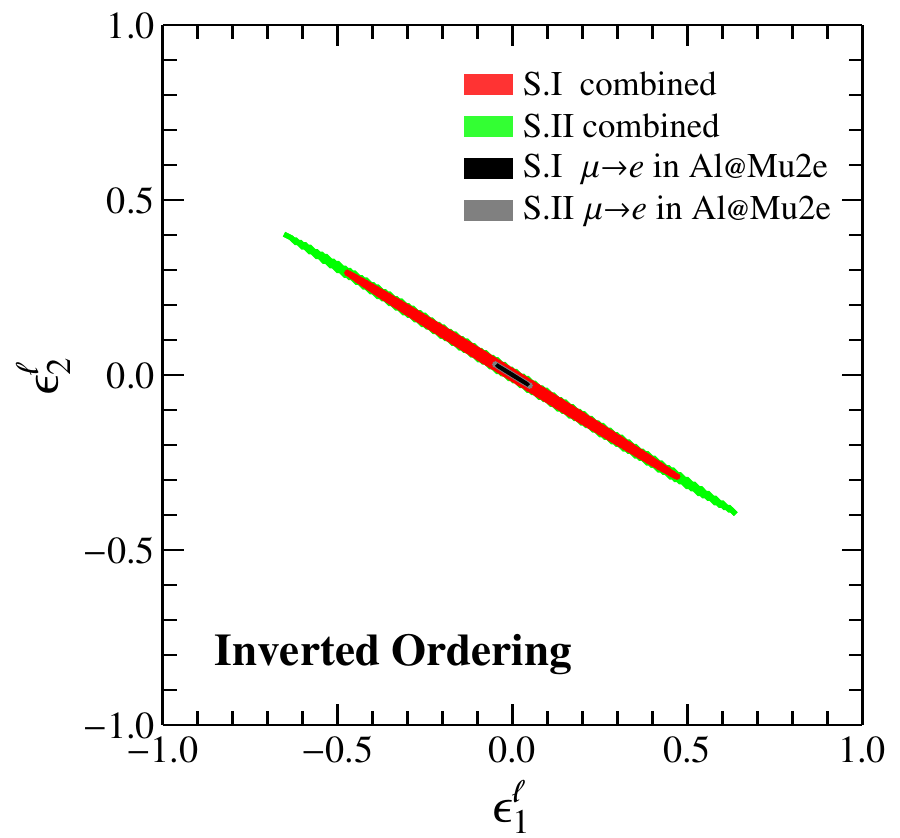}
  \caption{Combined constraints on $(\eu_0,\,\ed_0,\,\el_0,\, \el_1, \,\el_2)$ at 90\% CL, plotted in the $(\el_1, \, \el_2)$ plane, in the NO (Left) and IO (Right) cases, which are dominated by the $\mu \to e\gamma$ decay. The red and green regions are the allowed parameter space in Scenario I and II, respectively. The tiny black and dark regions indicate the future sensitivity to the $\mu \to e$ conversion in Al at the Mu2e experiment.}
  \label{fig:el1:el2:combined}
\end{figure}

To constrain the MFV parameters, we consider various LFV processes including $h \to \ell_i \ell_j$, $\ell_i \to \ell_j \ell_k \bar\ell_l$, $\ell_i \to \ell_j \gamma$, $\mu \to e $ conversion in nuclei, leptonic EDM, and anomalous magnetic moment. The previously obtained bounds on $(\eu_0, \, \ed_0, \, \el_0)$ from the Higgs data have been also included. After combining all these constraints, the allowed parameter space of $(\eu_0,\,\ed_0,\,\el_0,\, \el_1, \,\el_2)$ are obtained for scenario I and II in the NO and IO cases, which are plotted in the $(\el_1,\el_2)$ plane in Fig.~\ref{fig:el1:el2:combined}. It is found that the most strong constraints on the MFV parameters come from the branching ratio of $\mu \to e \gamma$ decay. Our detailed numerical analysis shows that the $\mu \to e \gamma$ decay in the allowed parameter space is dominated by the two-loop contribution $c_R^{2-{\rm loop}}$ in eq.~\eqref{eq:two-loop}, which is proportional to the couplings $Y_R^{e\mu}$ and $Y_L^{tt}$. Due to the values of the PMNS matrix in the IO case, the contributions from $\el_1$ and $\el_2$ can strongly cancel to each other in the Yukawa couplings $Y_R^{e\mu}$. It makes the allowed ranges of $\el_1$ and $\el_2$ in the IO case are much wider than the one in the NO case but have larger fine-tuning.

\begin{figure}[t]
  \centering
  \includegraphics[width=0.4\textwidth]{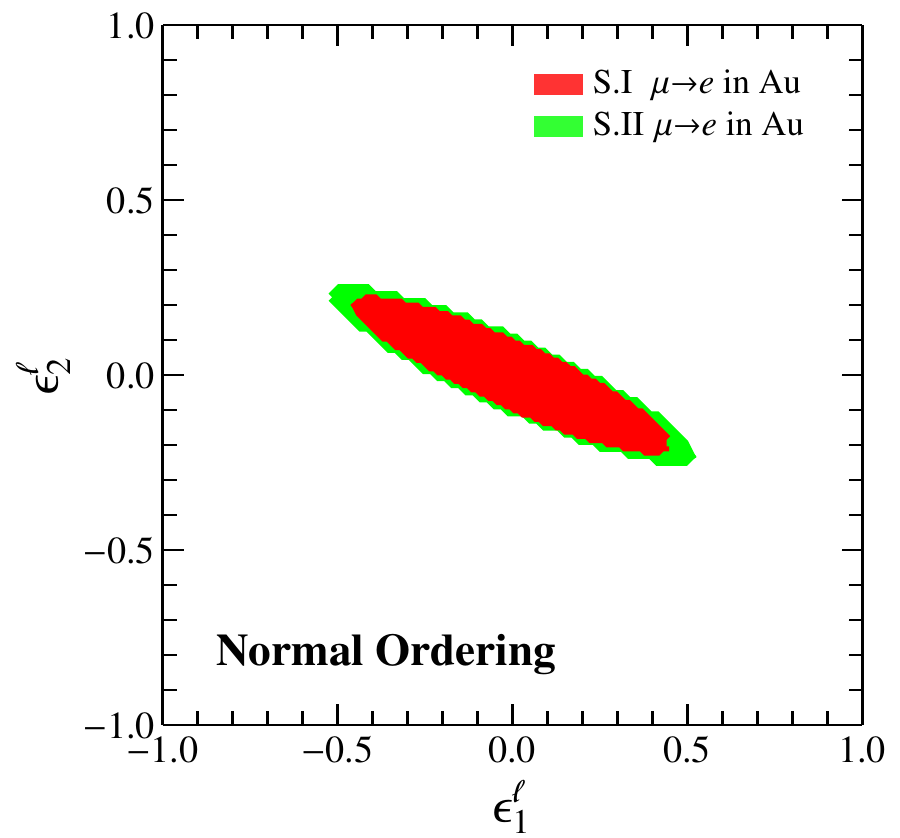}
  \qquad
  \includegraphics[width=0.4\textwidth]{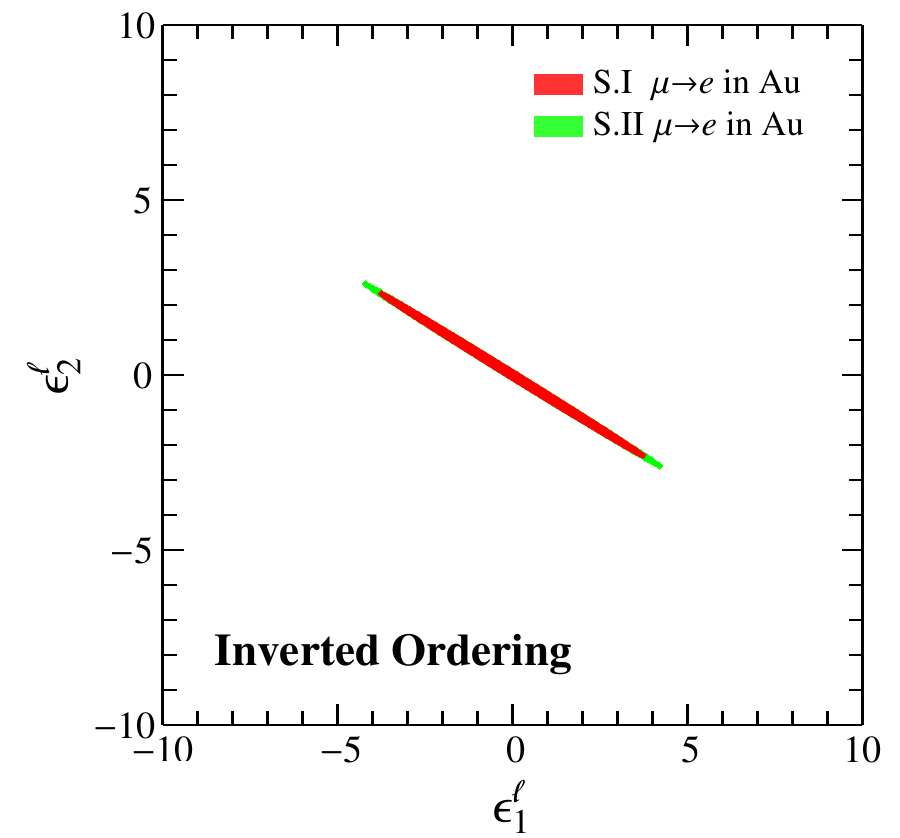}
  \caption{The same as Fig.~\ref{fig:el1:el2:combined}, but only under the constraint of $\mu \to e$ conversion in Au.}
  \label{fig:el1:el2:lm_le_conversion}
\end{figure}

For comparison, the bounds from $\mu \to e$ conversion in Au are shown in Fig.~\ref{fig:el1:el2:lm_le_conversion}, which are much weaker than ones from the $\mu \to e \gamma$ decay. In the future Mu2e experiment, the sensitivity for the branching ratio of $\mu \to e$ conversion is expected to be improved by 4 orders of magnitude compared to the current SINDRUM II bound, which corresponds to $7 \times 10^{-17}$ in Al at 90\% CL~\cite{Abusalma:2018xem}. The allowed parameter space corresponding to the future sensitivity at the Mu2e experiment are shown in Fig.~\ref{fig:el1:el2:combined}. It can be seen that the expected bounds at the Mu2e experiment are much more stringent than the ones obtained from the current measurements on $\mu \to e \gamma$ decay. In the near future, with three-year run, the MEG II experiment can reach a sensitivity of $6 \times 10^{-14}$ at 90\% CL for $\mB(\mu \to e \gamma)$~\cite{Baldini:2018nnn}. However, the corresponding bounds on the MFV parameters are much weaker than the ones experted at the Mu2e experiment.

\begin{figure}[t]
  \centering
  \includegraphics[width=0.4\textwidth]{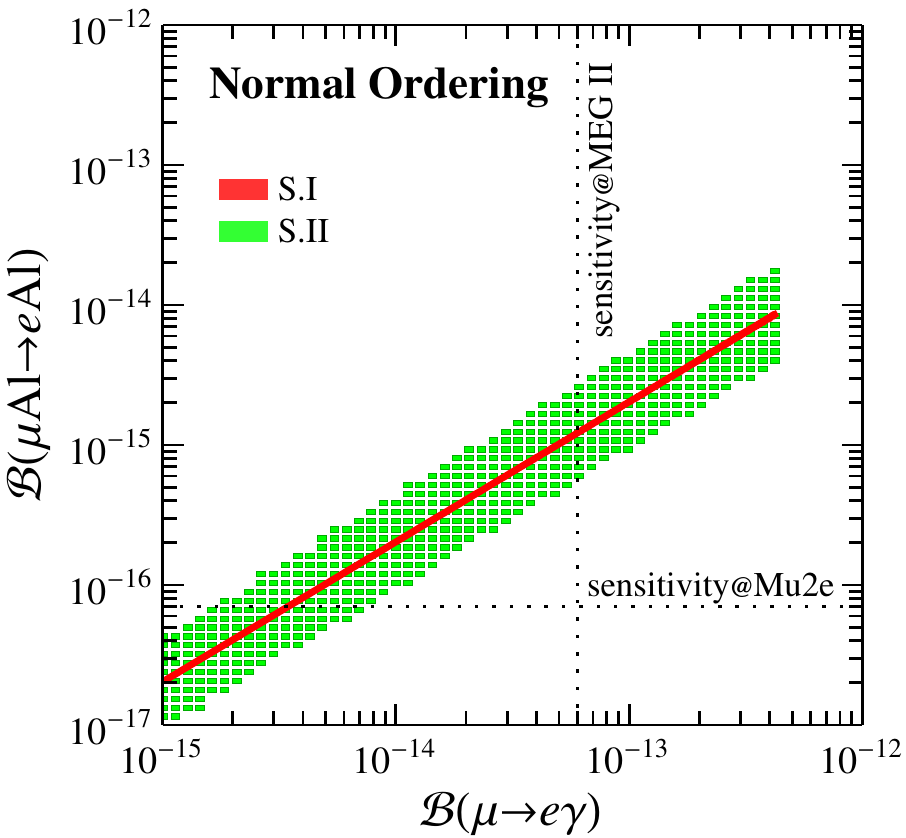}
  \qquad
  \includegraphics[width=0.4\textwidth]{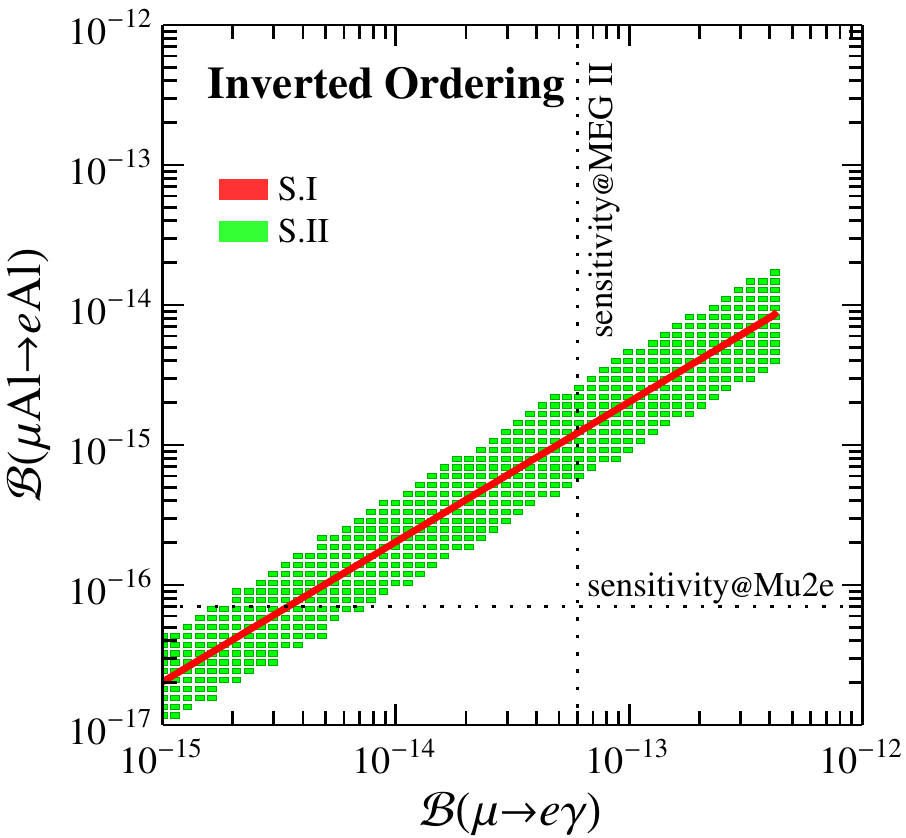}
  \caption{Correlation between $\mB(\mu \to e\gamma)$ and $\mB(\mu {\rm Au} \to e{\rm Au})$, in the NO (left) and IO (right) cases. The red and black region denotes the S.I and S.II, respectively.}
  \label{fig:correlation}
\end{figure}

Since the experimental sensitivity to the LFV processes $\mu \to e\gamma$ and $\mu \to e$ conversion in nuclei will be greatly improved in the near future, we show the correlations between $\mB(\mu \to e\gamma)$ and $\mB(\mu \, {\rm Al} \to e \, {\rm Al})$ in Fig.~\ref{fig:correlation}, which are obtained in the allowed parameter space corresponding to Fig.~\ref{fig:el1:el2:combined}. It can be seen that, the correlations in the NO and IO cases are almost the same. To understand this, we should notice that the Higgs FCNC effects on these two processes are dominated by the contributions $c_R^{2-{\rm loop}}$ and $g_{LS}^q$ in the allowed parameter space in both the NO and IO cases. In the scenario I, from their definitions in eq.~\eqref{eq:two-loop} and \eqref{eq:WC:scalar}, they are proportional to the Yukawa coupling $Y_R^{e\mu}$, which makes the branching ratios of both the two processes are proportional to $|Y_R^{e\mu}|^2$. Therefore, although $Y_R^{e\mu}$ depends on $(\el_1,\,\el_2)$ differently in the NO and IO cases, the correlation between $\mB(\mu\,{\rm Al} \to e \,{\rm Al})$ and $\mB(\mu \to e\gamma)$ is very strong and does no depend on the ordering of the light neutrinos' masses, as shown by the thin red regions in Fig.~\ref{fig:correlation}. In the scenario II, the contributions $c_R^{2-{\rm loop}}$ and $g_{LS}^q$ are also proportional to $Y_L^{tt}$ and $Y_R^{qq}$, respectively. In the MFV framework, the flavor-conserving couplings $Y_L^{tt}$ and $Y_R^{qq}$ mainly depend on the parameters $(\eu_0,\,\ed_0)$ and their dependence are the same between in the NO and IO cases. These flavor-conserving couplings make the correlation between $\mB(\mu\,{\rm Al} \to e \,{\rm Al})$ and $\mB(\mu \to e\gamma)$ much weaker than the one in the scenarion I in both the NO and IO cases, as shown by the wide green regions in Fig.~\ref{fig:correlation}. Considering the bounds on the flavor-conserving couplings will be largely improved by the future LHC data, the correlation in the scenario II is expected to become much stronger and approach the one in the scenario I.

For the anomalous magnetic moment $a_\mu$, current data show about $3\sigma$ deviation from the SM prediction~\cite{Lindner:2016bgg,PDG:2018}. In the MFV framework, explanation for this anomaly needs large LFV parameters $\el_1$ and $\el_2$, which is ruled out by the $\mu \to e \gamma$ decay.

Using the combined bounds obtained in the previous sections, the upper limits on various LFV $B_s$ and Higgs decays are obtained for the scenario I and II and in the NO and IO cases, which are shown in Tab.~\ref{tab:prediction}. For the $h \to \mu \tau$ decay,  the upper limits in the MFV are about two orders of magnitude lower than the current LHC bounds, which make searches for this channel challenging at the LHC. For the other LFV decays, since the upper bounds on their branching ratios are lower than the current LHC bounds by several orders of magnitude, they are very difficult to be measured at the LHC. For the $B_s \to \mu^+ \mu^-$ decay in both the NO and IO cases, it is found that its branching ratio can't deviate from the SM prediction by more than 1\%.

\begin{table}[t]
  \centering
  \begin{tabular}{|r | l |c | c | c | c | c | c |}
    \hline
   & &$\Gamma(h \to e\mu)$ & $\Gamma(h \to e \tau)$ & $\Gamma(h \to \mu\tau)$ & $\mB(B_s \to e \mu)$ & $\mB(B_s \to e \tau)$ & $\mB(B_s \to \mu \tau)$ 
   \\\hline
   NO & S.I 
   & 
   $1.2\times 10^{-8}$ & $1.3 \times 10^{-5}$ & $9.0\times 10^{-5}$ 
   &
   $2.4 \times 10^{-16}$ & $2.6\times 10^{-13}$ & $1.8 \times 10^{-12}$ 
   \\\hline
   NO &S.II 
   & 
   $2.2 \times 10^{-8}$ & $2.4 \times 10^{-5}$ & $1.7 \times 10^{-4}$ 
   &
   $4.6 \times 10^{-16}$ & $5.0 \times 10^{-13}$ & $3.5 \times 10^{-12}$
   \\\hline
   IO & S.I 
   & 
   $1.2 \times 10^{-8}$ & $4.7 \times 10^{-6}$ & $7.1 \times 10^{-5}$ 
   &
   $2.4 \times 10^{-16}$ & $9.6 \times 10^{-14}$ & $1.4 \times 10^{-12}$
   \\\hline
   IO & S.II 
   & 
   $2.2 \times 10^{-8}$ & $8.7 \times 10^{-6}$ & $1.3 \times 10^{-4}$ 
   &
   $4.5 \times 10^{-16}$ & $1.8 \times 10^{-13}$ & $2.6 \times 10^{-12}$
    \\\hline
  \end{tabular}
  \caption{Upper bounds on $\Gamma(h \to \ell_i \ell_j)$ [MeV] and $\mB(B_s \to \ell_i \ell_j)$ at 90\%CL.}
  \label{tab:prediction}
\end{table}

\section{Conclusions}\label{sec:conclusion}
Motivated by the recent LHC searches for the LFV decays $B_s\to \ell_i\ell_j$ and $h\to \ell_i \ell_j$, we study the tree-level Higgs FCNC interactions in the EFT approach. With and without the MFV hypothesis, we investigate the Higgs FCNC effects on the $B_s - \bar B_s$, $B_d -\bar B_d$ and $K^0 - \bar K^0$ mixing, the lepton FCNC processes $\ell_i \to \ell_j \gamma$, $\ell_i \to \ell_j\ell_k \bar \ell_l$, $\mu \to e$ conversion in nuclei, the LHC Higgs data, and etc, and derive the bounds on the Higgs FCNC couplings.

In the general case, the two LFV decays $B_s \to \ell_1 \ell_2$ and $h\to \ell_1 \ell_2$ are related to each other by the following expression
\begin{align*}
  \frac{\mB(B_s \to \ell_1 \ell_2)}{\mB(h \to \ell_1 \ell_2)} \approx 2.1 |\bar Y_{sb}|^2,
\end{align*}
assuming the SM Higgs total width. After deriving the bounds on $\bar Y_{sb}$ from $B_s - \bar B_s$ mixing and $B_s \to \mu^+ \mu^-$, predictions on various Higgs and $B_s$ FCNC decays are obtained, such as
\begin{align*}
  \mB(B_s \to e \mu) <2.1 \times 10^{-9}, \qquad \mB(h \to sb) < 4.1 \times 10^{-2},
\end{align*}
at 95\% CL, where the SM Higgs total width is assumed.

In the MFV hypothesis, strong constraints on the free parameters $(\eu_0,\,\ed_0,\,\ed_1,\,\el_0,\,\el_1,\,\el_2)$ are derived. We find that the bounds on $(\eu_0,\,\ed_0,\,\el_0)$ are dominated by the LHC Higgs data, $\ed_1$ the $B_s - \bar B_s$ mixing, and $(\el_1,\,\el_2)$ the $\mu \to e \gamma$ decay. Using these constraints, we obtain upper limits on various FCNC processes, such as
\begin{align*}
  \mB(h \to sb) < 4.9 \times 10^{-4},  
\end{align*}
at 95\% CL, and for the normal (inverted) ordering of the light neutrinos' masses, 
\begin{align*}
  \mB(h \to \mu \tau) < 4.2\,(3.2)\times 10^{-5},
  \qquad
  \mB(B_s\to e\mu)<4.6\,(4.5)\times 10^{-16},
\end{align*}
at 90\% CL, where the SM Higgs total width is assumed. For the $B_s \to \mu^+ \mu^-$ decay, its branching ratio can't deviate from the SM prediction by more than 1\%. For the various $B_s \to \ell_1 \ell_2$ and $h \to \ell_1 \ell_2$ decays, since the upper limits of their branching ratios are much lower than the current LHC bounds, searches for these LFV processes are very challenging at the LHC. However, with the improved measurements at the future MEG II and Mu2e experiments, searches for the LFV Higgs couplings in the $\mu \to e \gamma$ decay and $\mu \to e$ conversion in Al are very promising. In the MFV, the branching ratios of these two processes are strongly correlated to each other. Our bounds and correlations for the various processes can be used to obtain valuable information about the Higgs FCNC couplings from future measurements at the LHC and the low-energy experiments.

\section*{Acknowledgments}
This work was supported in part by the MOST (Grant No. MOST 106-2112-M-002-003-MY3 ), and in part by Key Laboratory for Particle Physics, Astrophysics and Cosmology, Ministry of Education, and Shanghai Key Laboratory for Particle Physics and Cosmology (Grant No. 15DZ2272100), and in part by the NSFC (Grant Nos. 11575111 and 11735010). XY thanks CCNU for its hospitality, where this work was partly conducted.

\bibliographystyle{JHEP}
\bibliography{ref}

\providecommand{\href}[2]{#2}\begingroup\raggedright\begin{thebibliography}{10}

\bibitem{Aad:2012tfa}
{\bf ATLAS} Collaboration, G.~Aad et~al., {\it {Observation of a new particle
  in the search for the Standard Model Higgs boson with the ATLAS detector at
  the LHC}},  {\em Phys. Lett.} {\bf B716} (2012) 1--29,
  [\href{http://arxiv.org/abs/1207.7214}{{\tt arXiv:1207.7214}}].

\bibitem{Chatrchyan:2012xdj}
{\bf CMS} Collaboration, S.~Chatrchyan et~al., {\it {Observation of a new boson
  at a mass of 125 GeV with the CMS experiment at the LHC}},  {\em Phys. Lett.}
  {\bf B716} (2012) 30--61, [\href{http://arxiv.org/abs/1207.7235}{{\tt
  arXiv:1207.7235}}].

\bibitem{Csaki:2015hcd}
C.~Csaki, C.~Grojean, and J.~Terning, {\it {Alternatives to an Elementary
  Higgs}},  {\em Rev. Mod. Phys.} {\bf 88} (2016), no.~4 045001,
  [\href{http://arxiv.org/abs/1512.00468}{{\tt arXiv:1512.00468}}].

\bibitem{Mariotti:2016owy}
C.~Mariotti and G.~Passarino, {\it {Higgs boson couplings: measurements and
  theoretical interpretation}},  {\em Int. J. Mod. Phys.} {\bf A32} (2017),
  no.~04 1730003, [\href{http://arxiv.org/abs/1612.00269}{{\tt
  arXiv:1612.00269}}].

\bibitem{Branco:2011iw}
G.~C. Branco, P.~M. Ferreira, L.~Lavoura, M.~N. Rebelo, M.~Sher, and J.~P.
  Silva, {\it {Theory and phenomenology of two-Higgs-doublet models}},  {\em
  Phys. Rept.} {\bf 516} (2012) 1--102,
  [\href{http://arxiv.org/abs/1106.0034}{{\tt arXiv:1106.0034}}].

\bibitem{Chiang:2009kb}
C.-W. Chiang, N.~G. Deshpande, X.-G. He, and J.~Jiang, {\it {The Family
  $SU(2)_l \times SU(2)_h \times U(1)$ Model}},  {\em Phys. Rev.} {\bf D81}
  (2010) 015006, [\href{http://arxiv.org/abs/0911.1480}{{\tt
  arXiv:0911.1480}}].

\bibitem{He:2002ha}
X.-G. He and G.~Valencia, {\it {The $Z \to b \bar{b}$ decay asymmetry and
  left-right models}},  {\em Phys. Rev.} {\bf D66} (2002) 013004,
  [\href{http://arxiv.org/abs/hep-ph/0203036}{{\tt hep-ph/0203036}}]. [Erratum:
  Phys. Rev.D66,079901(2002)].

\bibitem{Crivellin:2013wna}
A.~Crivellin, A.~Kokulu, and C.~Greub, {\it {Flavor-phenomenology of
  two-Higgs-doublet models with generic Yukawa structure}},  {\em Phys. Rev.}
  {\bf D87} (2013), no.~9 094031, [\href{http://arxiv.org/abs/1303.5877}{{\tt
  arXiv:1303.5877}}].

\bibitem{Kim:2015zla}
C.~S. Kim, Y.~W. Yoon, and X.-B. Yuan, {\it {Exploring top quark FCNC within
  2HDM type III in association with flavor physics}},  {\em JHEP} {\bf 12}
  (2015) 038, [\href{http://arxiv.org/abs/1509.00491}{{\tt arXiv:1509.00491}}].

\bibitem{Aaij:2017cza}
{\bf LHCb} Collaboration, R.~Aaij et~al., {\it {Search for the lepton-flavour
  violating decays B$_{(s)}^{0} \to e^{\pm}\mu^{\mp}$}},  {\em JHEP} {\bf 03}
  (2018) 078, [\href{http://arxiv.org/abs/1710.04111}{{\tt arXiv:1710.04111}}].

\bibitem{Sirunyan:2017xzt}
{\bf CMS} Collaboration, A.~M. Sirunyan et~al., {\it {Search for lepton flavour
  violating decays of the Higgs boson to $\mu\tau$ and e$\tau$ in proton-proton
  collisions at $\sqrt{s}=$ 13 TeV}},  {\em Submitted to: JHEP} (2017)
  [\href{http://arxiv.org/abs/1712.07173}{{\tt arXiv:1712.07173}}].

\bibitem{Khachatryan:2016rke}
{\bf CMS} Collaboration, V.~Khachatryan et~al., {\it {Search for lepton flavour
  violating decays of the Higgs boson to $e \tau$ and $e \mu$ in
  proton–proton collisions at $\sqrt s=$ 8 TeV}},  {\em Phys. Lett.} {\bf
  B763} (2016) 472--500, [\href{http://arxiv.org/abs/1607.03561}{{\tt
  arXiv:1607.03561}}].

\bibitem{Khachatryan:2015kon}
{\bf CMS} Collaboration, V.~Khachatryan et~al., {\it {Search for
  Lepton-Flavour-Violating Decays of the Higgs Boson}},  {\em Phys. Lett.} {\bf
  B749} (2015) 337--362, [\href{http://arxiv.org/abs/1502.07400}{{\tt
  arXiv:1502.07400}}].

\bibitem{He:2015rqa}
X.-G. He, J.~Tandean, and Y.-J. Zheng, {\it {Higgs decay h → μτ with
  minimal flavor violation}},  {\em JHEP} {\bf 09} (2015) 093,
  [\href{http://arxiv.org/abs/1507.02673}{{\tt arXiv:1507.02673}}].

\bibitem{Abusalma:2018xem}
{\bf Mu2e} Collaboration, F.~Abusalma et~al., {\it {Expression of Interest for
  Evolution of the Mu2e Experiment}},
  \href{http://arxiv.org/abs/1802.02599}{{\tt arXiv:1802.02599}}.

\bibitem{Aaij:2014ora}
{\bf LHCb} Collaboration, R.~Aaij et~al., {\it {Test of lepton universality
  using $B^{+}\rightarrow K^{+}\ell^{+}\ell^{-}$ decays}},  {\em Phys. Rev.
  Lett.} {\bf 113} (2014) 151601, [\href{http://arxiv.org/abs/1406.6482}{{\tt
  arXiv:1406.6482}}].

\bibitem{Aaij:2017vbb}
{\bf LHCb} Collaboration, R.~Aaij et~al., {\it {Test of lepton universality
  with $B^{0} \rightarrow K^{*0}\ell^{+}\ell^{-}$ decays}},  {\em JHEP} {\bf
  08} (2017) 055, [\href{http://arxiv.org/abs/1705.05802}{{\tt
  arXiv:1705.05802}}].

\bibitem{Amhis:2014hma}
{\bf Heavy Flavor Averaging Group (HFAG)} Collaboration, Y.~Amhis et~al., {\it
  {Averages of $b$-hadron, $c$-hadron, and $\tau$-lepton properties as of
  summer 2014}},  \href{http://arxiv.org/abs/1412.7515}{{\tt arXiv:1412.7515}}.

\bibitem{Chen:2013qta}
K.-F. Chen, W.-S. Hou, C.~Kao, and M.~Kohda, {\it {When the Higgs meets the
  Top: Search for $t \to c h^0$ at the LHC}},  {\em Phys. Lett.} {\bf B725}
  (2013) 378--381, [\href{http://arxiv.org/abs/1304.8037}{{\tt
  arXiv:1304.8037}}].

\bibitem{Harnik:2012pb}
R.~Harnik, J.~Kopp, and J.~Zupan, {\it {Flavor Violating Higgs Decays}},  {\em
  JHEP} {\bf 03} (2013) 026, [\href{http://arxiv.org/abs/1209.1397}{{\tt
  arXiv:1209.1397}}].

\bibitem{Chivukula:1987py}
R.~S. Chivukula and H.~Georgi, {\it {Composite Technicolor Standard Model}},
  {\em Phys. Lett.} {\bf B188} (1987) 99--104.

\bibitem{Buras:2000dm}
A.~J. Buras, P.~Gambino, M.~Gorbahn, S.~Jager, and L.~Silvestrini, {\it
  {Universal unitarity triangle and physics beyond the standard model}},  {\em
  Phys. Lett.} {\bf B500} (2001) 161--167,
  [\href{http://arxiv.org/abs/hep-ph/0007085}{{\tt hep-ph/0007085}}].

\bibitem{DAmbrosio:2002vsn}
G.~D'Ambrosio, G.~F. Giudice, G.~Isidori, and A.~Strumia, {\it {Minimal flavor
  violation: An Effective field theory approach}},  {\em Nucl. Phys.} {\bf
  B645} (2002) 155--187, [\href{http://arxiv.org/abs/hep-ph/0207036}{{\tt
  hep-ph/0207036}}].

\bibitem{Grzadkowski:2010es}
B.~Grzadkowski, M.~Iskrzynski, M.~Misiak, and J.~Rosiek, {\it {Dimension-Six
  Terms in the Standard Model Lagrangian}},  {\em JHEP} {\bf 10} (2010) 085,
  [\href{http://arxiv.org/abs/1008.4884}{{\tt arXiv:1008.4884}}].

\bibitem{Chiang:2017etj}
C.-W. Chiang, X.-G. He, F.~Ye, and X.-B. Yuan, {\it {Constraints and
  Implications on Higgs FCNC Couplings from Precision Measurement of $B_s \to
  \mu^+\mu^-$ Decay}},  {\em Phys. Rev.} {\bf D96} (2017), no.~3 035032,
  [\href{http://arxiv.org/abs/1703.06289}{{\tt arXiv:1703.06289}}].

\bibitem{Colangelo:2008qp}
G.~Colangelo, E.~Nikolidakis, and C.~Smith, {\it {Supersymmetric models with
  minimal flavour violation and their running}},  {\em Eur. Phys. J.} {\bf C59}
  (2009) 75--98, [\href{http://arxiv.org/abs/0807.0801}{{\tt
  arXiv:0807.0801}}].

\bibitem{Mercolli:2009ns}
L.~Mercolli and C.~Smith, {\it {EDM constraints on flavored CP-violating
  phases}},  {\em Nucl. Phys.} {\bf B817} (2009) 1--24,
  [\href{http://arxiv.org/abs/0902.1949}{{\tt arXiv:0902.1949}}].

\bibitem{Chiang:2017hlj}
C.-W. Chiang, X.-G. He, J.~Tandean, and X.-B. Yuan, {\it {$R_{K^{(*)}}$ and
  related $b\to s\ell\bar\ell$ anomalies in minimal flavor violation framework
  with $Z'$ boson}},  {\em Phys. Rev.} {\bf D96} (2017), no.~11 115022,
  [\href{http://arxiv.org/abs/1706.02696}{{\tt arXiv:1706.02696}}].

\bibitem{He:2014uya}
X.-G. He, C.-J. Lee, S.-F. Li, and J.~Tandean, {\it {Fermion EDMs with Minimal
  Flavor Violation}},  {\em JHEP} {\bf 08} (2014) 019,
  [\href{http://arxiv.org/abs/1404.4436}{{\tt arXiv:1404.4436}}].

\bibitem{He:2014efa}
X.-G. He, C.-J. Lee, J.~Tandean, and Y.-J. Zheng, {\it {Seesaw Models with
  Minimal Flavor Violation}},  {\em Phys. Rev.} {\bf D91} (2015), no.~7 076008,
  [\href{http://arxiv.org/abs/1411.6612}{{\tt arXiv:1411.6612}}].

\bibitem{He:2014fva}
X.-G. He, C.-J. Lee, S.-F. Li, and J.~Tandean, {\it {Large electron electric
  dipole moment in minimal flavor violation framework with Majorana
  neutrinos}},  {\em Phys. Rev.} {\bf D89} (2014), no.~9 091901,
  [\href{http://arxiv.org/abs/1401.2615}{{\tt arXiv:1401.2615}}].

\bibitem{Cirigliano:2005ck}
V.~Cirigliano, B.~Grinstein, G.~Isidori, and M.~B. Wise, {\it {Minimal flavor
  violation in the lepton sector}},  {\em Nucl. Phys.} {\bf B728} (2005)
  121--134, [\href{http://arxiv.org/abs/hep-ph/0507001}{{\tt hep-ph/0507001}}].

\bibitem{Cirigliano:2006su}
V.~Cirigliano and B.~Grinstein, {\it {Phenomenology of minimal lepton flavor
  violation}},  {\em Nucl. Phys.} {\bf B752} (2006) 18--39,
  [\href{http://arxiv.org/abs/hep-ph/0601111}{{\tt hep-ph/0601111}}].

\bibitem{Alonso:2011jd}
R.~Alonso, G.~Isidori, L.~Merlo, L.~A. Munoz, and E.~Nardi, {\it {Minimal
  flavour violation extensions of the seesaw}},  {\em JHEP} {\bf 06} (2011)
  037, [\href{http://arxiv.org/abs/1103.5461}{{\tt arXiv:1103.5461}}].

\bibitem{Dinh:2017smk}
D.~N. Dinh, L.~Merlo, S.~T. Petcov, and R.~Vega-Álvarez, {\it {Revisiting
  Minimal Lepton Flavour Violation in the Light of Leptonic CP Violation}},
  {\em JHEP} {\bf 07} (2017) 089, [\href{http://arxiv.org/abs/1705.09284}{{\tt
  arXiv:1705.09284}}].

\bibitem{Casas:2001sr}
J.~A. Casas and A.~Ibarra, {\it {Oscillating neutrinos and muon ---> e,
  gamma}},  {\em Nucl. Phys.} {\bf B618} (2001) 171--204,
  [\href{http://arxiv.org/abs/hep-ph/0103065}{{\tt hep-ph/0103065}}].

\bibitem{Buras:2001ra}
A.~J. Buras, S.~Jager, and J.~Urban, {\it {Master formulae for $\Delta F=2$ NLO
  QCD factors in the standard model and beyond}},  {\em Nucl. Phys.} {\bf B605}
  (2001) 600--624, [\href{http://arxiv.org/abs/hep-ph/0102316}{{\tt
  hep-ph/0102316}}].

\bibitem{Buchalla:1995vs}
G.~Buchalla, A.~J. Buras, and M.~E. Lautenbacher, {\it {Weak decays beyond
  leading logarithms}},  {\em Rev. Mod. Phys.} {\bf 68} (1996) 1125--1144,
  [\href{http://arxiv.org/abs/hep-ph/9512380}{{\tt hep-ph/9512380}}].

\bibitem{Carrasco:2013zta}
{\bf ETM} Collaboration, N.~Carrasco et~al., {\it {B-physics from $N_f$ = 2
  tmQCD: the Standard Model and beyond}},  {\em JHEP} {\bf 03} (2014) 016,
  [\href{http://arxiv.org/abs/1308.1851}{{\tt arXiv:1308.1851}}].

\bibitem{Bazavov:2016nty}
{\bf Fermilab Lattice, MILC} Collaboration, A.~Bazavov et~al., {\it
  {$B^0_{(s)}$-mixing matrix elements from lattice QCD for the Standard Model
  and beyond}},  {\em Phys. Rev.} {\bf D93} (2016), no.~11 113016,
  [\href{http://arxiv.org/abs/1602.03560}{{\tt arXiv:1602.03560}}].

\bibitem{Artuso:2015swg}
M.~Artuso, G.~Borissov, and A.~Lenz, {\it {CP violation in the $B_s^0$
  system}},  {\em Rev. Mod. Phys.} {\bf 88} (2016), no.~4 045002,
  [\href{http://arxiv.org/abs/1511.09466}{{\tt arXiv:1511.09466}}].

\bibitem{Buchalla:1993bv}
G.~Buchalla and A.~J. Buras, {\it {QCD corrections to rare K and B decays for
  arbitrary top quark mass}},  {\em Nucl. Phys.} {\bf B400} (1993) 225--239.

\bibitem{Misiak:1999yg}
M.~Misiak and J.~Urban, {\it {QCD corrections to FCNC decays mediated by Z
  penguins and W boxes}},  {\em Phys. Lett.} {\bf B451} (1999) 161--169,
  [\href{http://arxiv.org/abs/hep-ph/9901278}{{\tt hep-ph/9901278}}].

\bibitem{Buchalla:1998ba}
G.~Buchalla and A.~J. Buras, {\it {The rare decays $K\to \pi \nu\bar\nu$, $B
  \to X \nu\bar\nu$ and $B \to l^+ l^-$: An Update}},  {\em Nucl. Phys.} {\bf
  B548} (1999) 309--327, [\href{http://arxiv.org/abs/hep-ph/9901288}{{\tt
  hep-ph/9901288}}].

\bibitem{Bobeth:2013tba}
C.~Bobeth, M.~Gorbahn, and E.~Stamou, {\it {Electroweak Corrections to $B_{s,d}
  \to \ell^+ \ell^-$}},  {\em Phys. Rev.} {\bf D89} (2014), no.~3 034023,
  [\href{http://arxiv.org/abs/1311.1348}{{\tt arXiv:1311.1348}}].

\bibitem{Hermann:2013kca}
T.~Hermann, M.~Misiak, and M.~Steinhauser, {\it {Three-loop QCD corrections to
  $B_s \to \mu^+ \mu^-$}},  {\em JHEP} {\bf 12} (2013) 097,
  [\href{http://arxiv.org/abs/1311.1347}{{\tt arXiv:1311.1347}}].

\bibitem{Bobeth:2013uxa}
C.~Bobeth, M.~Gorbahn, T.~Hermann, M.~Misiak, E.~Stamou, and M.~Steinhauser,
  {\it {$B_{s,d} \to l^+ l^-$ in the Standard Model with Reduced Theoretical
  Uncertainty}},  {\em Phys. Rev. Lett.} {\bf 112} (2014) 101801,
  [\href{http://arxiv.org/abs/1311.0903}{{\tt arXiv:1311.0903}}].

\bibitem{Li:2014fea}
X.-Q. Li, J.~Lu, and A.~Pich, {\it {$B_{s,d}^0 \to \ell^+\ell^-$ Decays in the
  Aligned Two-Higgs-Doublet Model}},  {\em JHEP} {\bf 06} (2014) 022,
  [\href{http://arxiv.org/abs/1404.5865}{{\tt arXiv:1404.5865}}].

\bibitem{Cheng:2015yfu}
X.-D. Cheng, Y.-D. Yang, and X.-B. Yuan, {\it {Revisiting $B_s \to \mu^+\mu^-$
  in the two-Higgs doublet models with $Z_2$ symmetry}},  {\em Eur. Phys. J.}
  {\bf C76} (2016), no.~3 151, [\href{http://arxiv.org/abs/1511.01829}{{\tt
  arXiv:1511.01829}}].

\bibitem{DeBruyn:2012wk}
K.~De~Bruyn, R.~Fleischer, R.~Knegjens, P.~Koppenburg, M.~Merk, A.~Pellegrino,
  and N.~Tuning, {\it {Probing New Physics via the $B^0_s\to \mu^+\mu^-$
  Effective Lifetime}},  {\em Phys. Rev. Lett.} {\bf 109} (2012) 041801,
  [\href{http://arxiv.org/abs/1204.1737}{{\tt arXiv:1204.1737}}].

\bibitem{Buras:2013uqa}
A.~J. Buras, R.~Fleischer, J.~Girrbach, and R.~Knegjens, {\it {Probing New
  Physics with the $B_s \to \mu^+ \mu^-$ Time-Dependent Rate}},  {\em JHEP}
  {\bf 07} (2013) 77, [\href{http://arxiv.org/abs/1303.3820}{{\tt
  arXiv:1303.3820}}].

\bibitem{Chang:1993kw}
D.~Chang, W.~S. Hou, and W.-Y. Keung, {\it {Two loop contributions of flavor
  changing neutral Higgs bosons to mu ---> e gamma}},  {\em Phys. Rev.} {\bf
  D48} (1993) 217--224, [\href{http://arxiv.org/abs/hep-ph/9302267}{{\tt
  hep-ph/9302267}}].

\bibitem{Kitano:2002mt}
R.~Kitano, M.~Koike, and Y.~Okada, {\it {Detailed calculation of lepton flavor
  violating muon electron conversion rate for various nuclei}},  {\em Phys.
  Rev.} {\bf D66} (2002) 096002,
  [\href{http://arxiv.org/abs/hep-ph/0203110}{{\tt hep-ph/0203110}}]. [Erratum:
  Phys. Rev.D76,059902(2007)].

\bibitem{Ellis:2008hf}
J.~R. Ellis, K.~A. Olive, and C.~Savage, {\it {Hadronic Uncertainties in the
  Elastic Scattering of Supersymmetric Dark Matter}},  {\em Phys. Rev.} {\bf
  D77} (2008) 065026, [\href{http://arxiv.org/abs/0801.3656}{{\tt
  arXiv:0801.3656}}].

\bibitem{Young:2009ps}
R.~D. Young and A.~W. Thomas, {\it {Recent results on nucleon sigma terms in
  lattice QCD}},  {\em Nucl. Phys.} {\bf A844} (2010) 266C--271C,
  [\href{http://arxiv.org/abs/0911.1757}{{\tt arXiv:0911.1757}}].

\bibitem{Suzuki:1987jf}
T.~Suzuki, D.~F. Measday, and J.~P. Roalsvig, {\it {Total Nuclear Capture Rates
  for Negative Muons}},  {\em Phys. Rev.} {\bf C35} (1987) 2212.

\bibitem{Jung:2012vu}
M.~Jung, X.-Q. Li, and A.~Pich, {\it {Exclusive radiative B-meson decays within
  the aligned two-Higgs-doublet model}},  {\em JHEP} {\bf 10} (2012) 063,
  [\href{http://arxiv.org/abs/1208.1251}{{\tt arXiv:1208.1251}}].

\bibitem{PDG:2018}
{\bf Particle Data Group} Collaboration, M.~Tanabashi et~al., {\it {Review of
  Particle Physics}},  {\em Phys. Rev.} {\bf D98} (2018) 030001.

\bibitem{Charles:2004jd}
{\bf CKMfitter Group} Collaboration, J.~Charles, A.~Hocker, H.~Lacker,
  S.~Laplace, F.~R. Le~Diberder, J.~Malcles, J.~Ocariz, M.~Pivk, and L.~Roos,
  {\it {CP violation and the CKM matrix: Assessing the impact of the asymmetric
  $B$ factories}},  {\em Eur. Phys. J.} {\bf C41} (2005), no.~1 1--131,
  [\href{http://arxiv.org/abs/hep-ph/0406184}{{\tt hep-ph/0406184}}].

\bibitem{Esteban:2016qun}
I.~Esteban, M.~C. Gonzalez-Garcia, M.~Maltoni, I.~Martinez-Soler, and
  T.~Schwetz, {\it {Updated fit to three neutrino mixing: exploring the
  accelerator-reactor complementarity}},  {\em JHEP} {\bf 01} (2017) 087,
  [\href{http://arxiv.org/abs/1611.01514}{{\tt arXiv:1611.01514}}].

\bibitem{Aoki:2016frl}
S.~Aoki et~al., {\it {Review of lattice results concerning low-energy particle
  physics}},  {\em Eur. Phys. J.} {\bf C77} (2017), no.~2 112,
  [\href{http://arxiv.org/abs/1607.00299}{{\tt arXiv:1607.00299}}].

\bibitem{Amhis:2016xyh}
{\bf HFLAV} Collaboration, Y.~Amhis et~al., {\it {Averages of $b$-hadron,
  $c$-hadron, and $\tau$-lepton properties as of summer 2016}},  {\em Eur.
  Phys. J.} {\bf C77} (2017), no.~12 895,
  [\href{http://arxiv.org/abs/1612.07233}{{\tt arXiv:1612.07233}}].

\bibitem{TheMEG:2016wtm}
{\bf MEG} Collaboration, A.~M. Baldini et~al., {\it {Search for the lepton
  flavour violating decay $\mu ^+ \rightarrow \mathrm {e}^+ \gamma $ with the
  full dataset of the MEG experiment}},  {\em Eur. Phys. J.} {\bf C76} (2016),
  no.~8 434, [\href{http://arxiv.org/abs/1605.05081}{{\tt arXiv:1605.05081}}].

\bibitem{Bertl:2006up}
{\bf SINDRUM II} Collaboration, W.~H. Bertl et~al., {\it {A Search for muon to
  electron conversion in muonic gold}},  {\em Eur. Phys. J.} {\bf C47} (2006)
  337--346.

\bibitem{Khachatryan:2016vau}
{\bf ATLAS, CMS} Collaboration, G.~Aad et~al., {\it {Measurements of the Higgs
  boson production and decay rates and constraints on its couplings from a
  combined ATLAS and CMS analysis of the LHC pp collision data at $ \sqrt{s}=7
  $ and 8 TeV}},  {\em JHEP} {\bf 08} (2016) 045,
  [\href{http://arxiv.org/abs/1606.02266}{{\tt arXiv:1606.02266}}].

\bibitem{Heinemeyer:2013tqa}
{\bf LHC Higgs Cross Section Working Group} Collaboration, J.~R. Andersen
  et~al., {\it {Handbook of LHC Higgs Cross Sections: 3. Higgs Properties}},
  \href{http://arxiv.org/abs/1307.1347}{{\tt arXiv:1307.1347}}.

\bibitem{Bernon:2015hsa}
J.~Bernon and B.~Dumont, {\it {Lilith: a tool for constraining new physics from
  Higgs measurements}},  {\em Eur. Phys. J.} {\bf C75} (2015), no.~9 440,
  [\href{http://arxiv.org/abs/1502.04138}{{\tt arXiv:1502.04138}}].

\bibitem{Aaltonen:2013ioz}
{\bf CDF, D0} Collaboration, T.~Aaltonen et~al., {\it {Higgs Boson Studies at
  the Tevatron}},  {\em Phys. Rev.} {\bf D88} (2013), no.~5 052014,
  [\href{http://arxiv.org/abs/1303.6346}{{\tt arXiv:1303.6346}}].

\bibitem{Buras:2013ooa}
A.~J. Buras and J.~Girrbach, {\it {Towards the Identification of New Physics
  through Quark Flavour Violating Processes}},  {\em Rept. Prog. Phys.} {\bf
  77} (2014) 086201, [\href{http://arxiv.org/abs/1306.3775}{{\tt
  arXiv:1306.3775}}].

\bibitem{Bertolini:2014sua}
S.~Bertolini, A.~Maiezza, and F.~Nesti, {\it {Present and Future K and B Meson
  Mixing Constraints on TeV Scale Left-Right Symmetry}},  {\em Phys. Rev.} {\bf
  D89} (2014), no.~9 095028, [\href{http://arxiv.org/abs/1403.7112}{{\tt
  arXiv:1403.7112}}].

\bibitem{Barducci:2017ioq}
D.~Barducci and A.~J. Helmboldt, {\it {Quark flavour-violating Higgs decays at
  the ILC}},  {\em JHEP} {\bf 12} (2017) 105,
  [\href{http://arxiv.org/abs/1710.06657}{{\tt arXiv:1710.06657}}].

\bibitem{Baldini:2018nnn}
{\bf MEG II} Collaboration, A.~M. Baldini et~al., {\it {The design of the MEG
  II experiment}},  \href{http://arxiv.org/abs/1801.04688}{{\tt
  arXiv:1801.04688}}.

\bibitem{Lindner:2016bgg}
M.~Lindner, M.~Platscher, and F.~S. Queiroz, {\it {A Call for New Physics : The
  Muon Anomalous Magnetic Moment and Lepton Flavor Violation}},  {\em Phys.
  Rept.} {\bf 731} (2018) 1--82, [\href{http://arxiv.org/abs/1610.06587}{{\tt
  arXiv:1610.06587}}].

\end{thebibliography}\endgroup

\end{document}